\pdfoutput=1 
\documentclass[a4paper,11pt]{article}

\usepackage{jinstpub} 

\usepackage{natbib} 

\title{\boldmath Fast neutron sensitivity of neutron detectors based on boron-10 converter layers }

\author[a,b,1]{G. Mauri,\note{Corresponding author.}}
\author[c,a]{F. Messi,}
\author[a]{K. Kanaki,}
\author[a,d]{R. Hall-Wilton,}
\author[a]{E. Karnickis,}
\author[a]{A. Khaplanov,}
\author[a]{F. Piscitelli}

\affiliation[a]{European Spallation Source ERIC (ESS), \\P.O. Box 176, SE-22100 Lund, Sweden.}
\affiliation[b]{Department of Physics, University of Perugia, \\Piazza Universit\'a 1, 06123 Perugia, Italy}
\affiliation[c]{Division of Nuclear Physics, Lund University, \\P.O. Box 118, SE-22100 Lund, Sweden.}
\affiliation[d]{Mid-Sweden University, \\SE-85170 Sundsvall, Sweden.}

\emailAdd{giacomo.mauri@esss.se}

\abstract{%Many neutron sources in the world are aiming to increase their performance and the European Spallation Source (ESS), under construction in Sweden, will be the brightest source in the world in the coming years. 
In the last few years many detector technologies for thermal neutron detection have been developed in order to face the shortage of $\mathrm{^3He}$, which is now much less available and more expensive. Moreover the $\mathrm{^3He}$-based detectors can not fulfil the requirements in performance, e.g. the spatial resolution and the counting rate capability needed for the new instruments. The Boron-10-based gaseous detectors have been proposed as a suitable choice. This and other alternatives technologies are being developed at ESS. Higher intensities mean higher signals but higher background as well. The signal-to-background ratio is an important feature to study, in particular the $\gamma$-ray and the fast neutron contributions. This paper investigates, for the first time, the fast neutrons sensitivity of $\mathrm{^{10}B}$-based thermal neutron detector. It presents the study of the detector response as a function of energy threshold and the underlying physical mechanisms. The latter are explained with the help of theoretical considerations and simulations.}


\begin{document}
\maketitle
\flushbottom

\section{Introduction}
The science progress in neutron physics are made possible thanks to the development both of sources and instrumentations, including neutron detectors. The European Spallation Source (ESS) will be a prominent infrastructure in this regard~\cite{ESS, ESS_TDR}. The ESS is designed to be the world's brightest neutron source. The instantaneous flux of the instruments will be the highest of any other existing neutron source.
\\ The $\mathrm{^3He}$-based techonologies have been the most used detectors for thermal neutrons. Both the availability and the requirements of higher performance are the reasons why a number of research programs are now aiming to find technologies that would replace the $\mathrm{^3He}$~\cite{HE3S_karl}. A promising technique is based on solid converter layers ($\mathrm{^{10}B}$, Gd) and gas proportional counter as the sensing medium. Some examples are the Multi-Grid~\cite{MG_2017,MG_IN6tests,MG_patent,MG_joni} for large area applications in neutron scattering, the Multi-Blade for neutron reflectometry~\cite{MIO_MB2014,MB_2017}, the Jalousie detector~\cite{DET_jalousie} for diffractometers, BandGEM~\cite{MPGD_GEMcroci, Bgem} and the Boron-coated straw-tubes~\cite{STRAW_lacy2011} for SANS and Gd-GEM~\cite{DET_doro1,gdgem} for neutron macromolecular diffractometer and CASCADE~\cite{DET_kohli}.
\\ The characterization of the background is a fundamental feature to study in order to understand the limit of the best signal-to-background ratio achievable. The main sources of background, which affect the detectors and instrument performance, are $\gamma$-rays, fast neutrons (1-10 MeV), environmental neutron counts (thermal and epithermal) and electronic noise. 
\\ In this paper we investigate the response of a Boron-10-based gaseous detector to fast neutrons. The characterization of the fast neutron sensitivity is important to define the best achievable signal-to-background ratio. A detailed description of the nuclear processes involved in the neutron interaction with the different materials of the detector is also reported. 
\\ All the measurements are performed at the Source Testing Facility of Lund University~\cite{IAEA-STF} with the Multi-Blade detector~\cite{MB_2017}, currently developed at ESS. A comparison with the $\gamma$-ray sensitivity~\cite{MG_gamma, MPGD_CrociGamma, MB_2017} follows the main analysis on fast neutron sensitivity.
\\ Theoretical analysis and Monte Carlo simulations are used to interpret the experimental results obtained on a Boron-10-based detector.
 
\section{Description of the setup}

\subsection{The Multi-Blade detector}\label{mbdescr}
 
\begin{figure}[htbp]
\centering
\includegraphics[width=.75\textwidth,keepaspectratio]{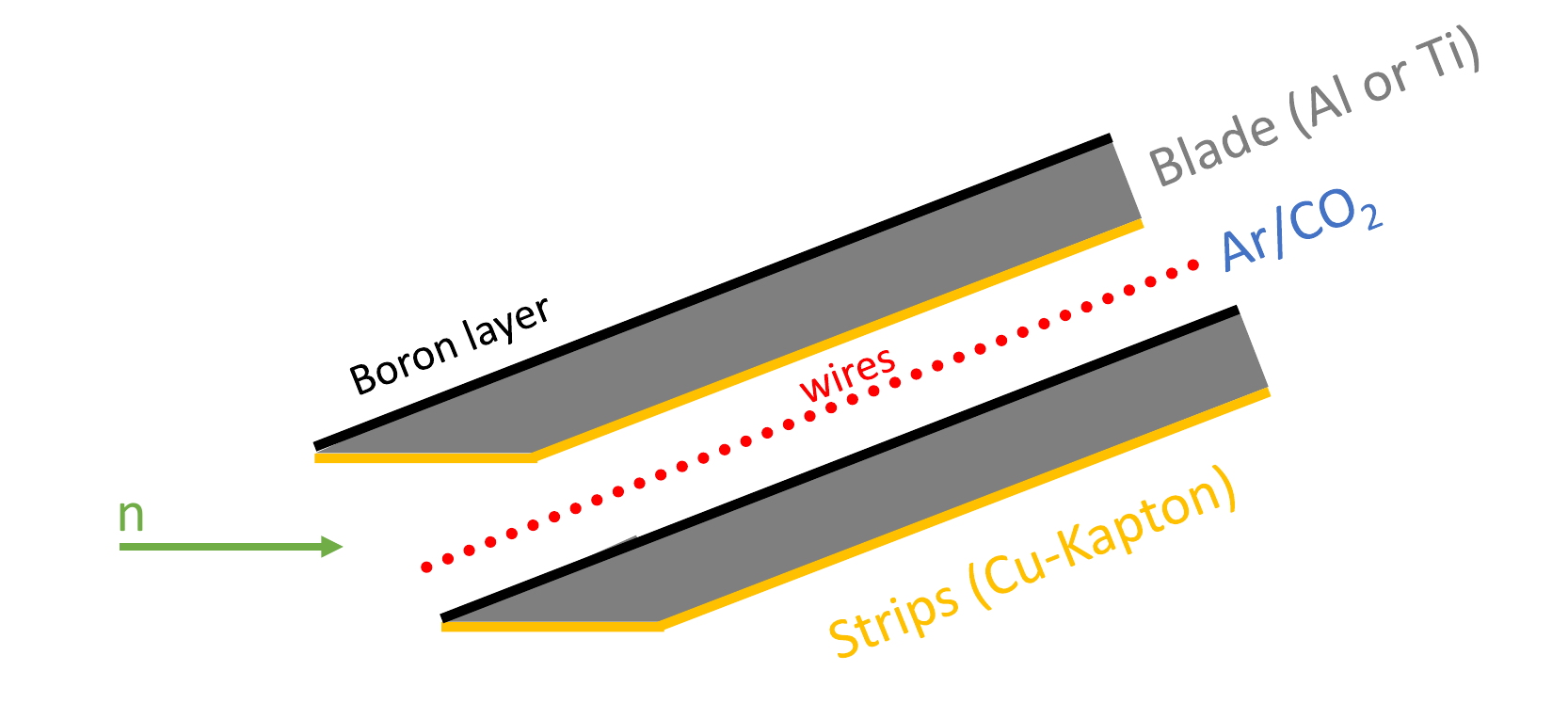}
\caption{\label{cstmb} \footnotesize Schematic view of the Multi-Blade cassettes.}
\end{figure}

The measurements presented in the paper have been performed with the Multi-Blade detector designed for neutron reflectometry applications. This detector is designed to improve the spatial resolution and the counting rate capability compared to present-day detectors. A detailed description of the detector can be found in \cite{MB_2017,MIO_MB2014}.
The Multi-Blade is a stack of Multi Wire Proportional Chambers (MWPC) operated at atmospheric pressure with a continuous gas flow (Ar/CO$_2$ 80/20 mixture). It is made up of identical units, called `cassettes'. Each cassette holds a `blade', a flat substrate of Aluminium or Titanium coated with $\mathrm{^{10}B_{4}C}$ \cite{B4C_carina,B4C_Schmidt}, and a two-dimensional readout system, which consists of a plane of wires and a plane of strips made of Copper on a substrate of polyimide (Kapton). The $\mathrm{^{10}B_{4}C}$ is the converter layer for thermal neutrons and the Ar/CO$_2$ is the sensing medium (a detailed description of the mechanism can be found in~\cite{MIO_analyt}). A sketch of the cassette is shown in figure~\ref{cstmb}. The materials in the detector that we consider in our study are: \textit{Ar , CO$_2$ , Al , Ti , Cu , Kapton , $^{10}$B$_{4}$C}.

\subsection{Setup} \label{subsetup}

\begin{figure}[htbp]
\centering
\includegraphics[width=1\textwidth,keepaspectratio]{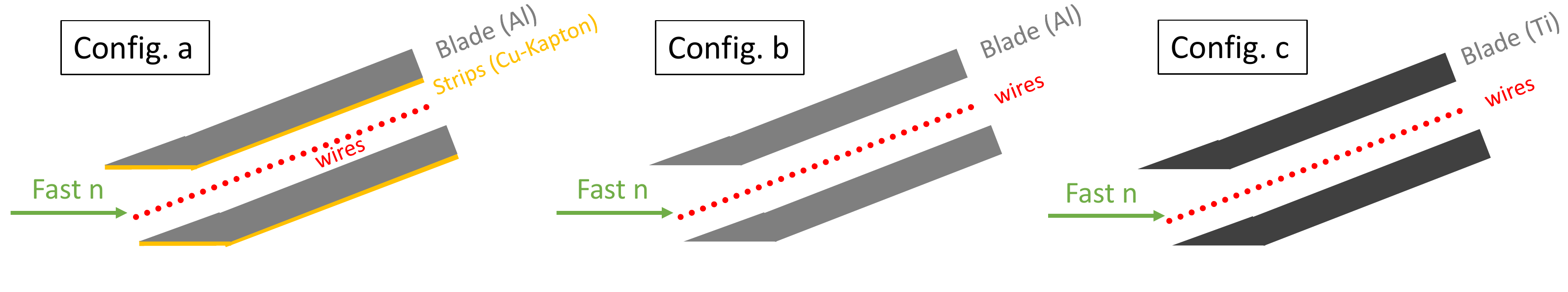}
\caption{\label{cassettesc} \footnotesize Schematic view of the Multi Blade cassettes for three different configurations. In all of them the $^{10}$B$_{4}$C layer was removed. a- Aluminium blade and strip-plane in Cu e Kapton, b- Aluminium blade and strip-plane removed, c- Titanium blade and strip-plane removed.}
\end{figure}

In order not to be sensitive to thermal neutrons, all the measurements were performed without the $^{10}$B layer, while keeping all the rest of the detector unchanged. We investigated three different detector configurations as shown in figure~\ref{cassettesc}. Config.(a) is the standard configuration which is a full cassette (without the $\mathrm{^{10}B_{4}C}$ layer see figure~\ref{cstmb} for comparison), Config.(b) is the configuration with Aluminum blades without strips, without Cu nor Kapton, and Config.(c) is the configuration with Titanium blades and no strips.  

\begin{figure}[htbp]
\centering
\includegraphics[width=1\textwidth,keepaspectratio]{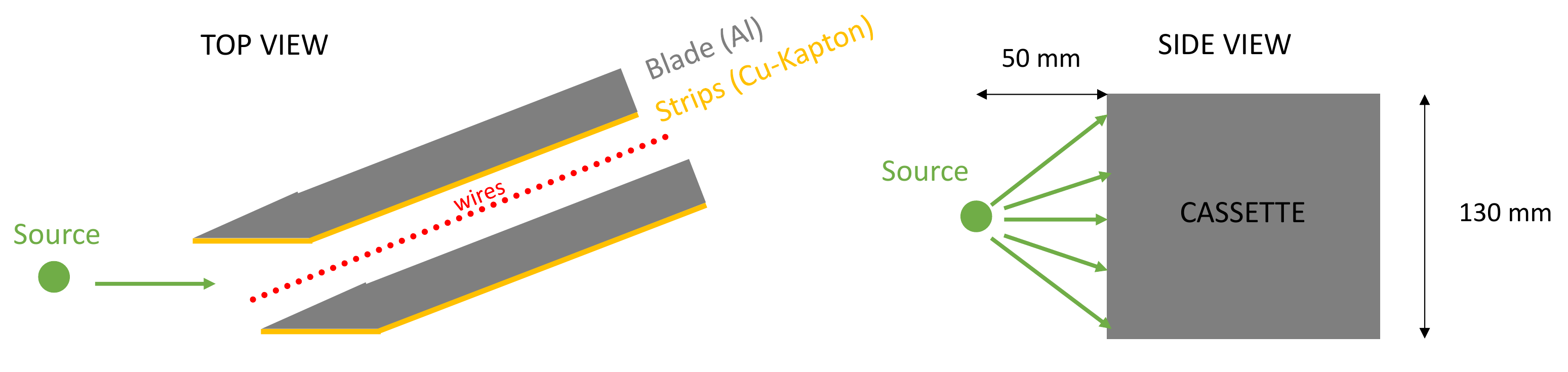}
\caption{\label{setup} \footnotesize Schematic representation of the experimental setup used to perform the measurements showing the position of the source.}
\end{figure}

All the measurements have been performed with the same experimental setup, a sketch of which is shown in figure~\ref{setup}. 

In the tests we measured the response of one cassette and this is representative of the full detector because it is modular and an event in a cassette does not influence neighbour cassettes. The cassette was equipped with an individual readout, one amplifier on each wire or strip (64 channels)~\cite{MB_2017}. The pre-amplifiers are CREMAT~\cite{EL_cremat} CR-110 charge sensitive pre-amplifiers and the signals are shaped with a CREMAT~\cite{EL_cremat} CR-200-500 ns Gaussian shaping amplifier. The pre-amplifiers have a gain of 1.4 V/pC and and the shaping amplifiers a gain of 10 with 500 ns shaping time. A CAEN~\cite{EL_CAEN} Digitizer is used to record the traces and the amplitude of the signals (Mod. V1740D, 12bit, 62.5 MS/s).
\\ A software threshold for each wire and strip is set to reject the electronic noise and  the detector is operated at 800 V (gas gain $\approx$15).

\subsection{Sources} \label{subsources}

Three neutron sources: $^{252}$Cf, $^{241}$Am/Be, $^{238}$Pu/Be, were used to perform the measurements. A further measurement was performed with a $^{60}$Co gamma source, in order to have a comparison with the previous work on gamma sensitivity~\cite{MG_gamma, MB_2017}.
\\Actinide/Be-based radioactive sources produce neutrons when the actinide decay by emission of an $\alpha$-particle which interacts with the $^9$Be nucleus. $^{252}$Cf decays by both $\alpha$-particle emission ($\sim$ 97\%) and spontaneous fission ($\sim$ 3\%), the latter giving rise to fast neutrons directly. The activity of the Actinide/Be sources used in this work is about one order of magnitude higher than that of the $^{252}$Cf.
\\The mean of the incident $\alpha$-particle energies is 5.48 MeV for the $^{241}$Am source and 5.49 MeV for the $^{238}$Pu source. The energy spectra of emitted neutrons are similar~\cite{PuBe_AmBe}. The fast neutron emission spectrum of $^{252}$Cf has an average energy of 2.1 MeV, with a most probable energy of 0.7 MeV~\cite{Cf}.
\\ Besides the differences between the energy spectra of the Actinide/Be-based radioactive sources and of the $^{252}$Cf, the energy distribution range considered is $1-10$ MeV for the three sources.
\\ We calculate the solid angle on a rectangular surface of the detector cassette under test (130x10)mm$^2$ for a fixed distance, d, between the source and the detector surface. We consider a point like source and we performed all the measurements with d = 50 mm, see figure~\ref{setup}. It results in a solid angle of $\approx 0.1$ sr ($\approx 1\%$ of a sphere). We estimate that the measurement of the distance between the source and the sensitive region can lead to an uncertainty which is less than a factor two variation for a misplacement of a $\pm 15$ mm from the fixed location of the source.
\\ We normalize all the measurements to the flux, number of neutrons per unit time and area, multiplying the activity by the solid angle. 

\section{Physical processes}\label{theocalc}

The thermal reaction mediated by a $\mathrm{^{10}B}$ layer is $\mathrm{^{10}B(n,\alpha)}$Li : 

\begin{equation}
n + ^{10}_{5}B \rightarrow \begin{cases}^{7}_{3}Li (1.02 MeV)+ \alpha ( 1.78 MeV) \quad 6\% \\ ^{7}_{3}Li (0.84 MeV)+ \alpha(1.47 MeV) + \gamma (0.48 MeV)\quad 94\% \end{cases}
\label{tras}
\end{equation}

The range of the product particles are of the order of a few $\mu$m in solids and a few mm in gasses at atmospheric pressure.

\subsection{Cross section and interaction probability calculation}\label{sp}

A preliminary study was performed on the nuclear processes that may occur between neutrons of energy range 1-10 MeV and nuclei present in the detector material. This investigation gives a qualitative indication which reactions are significant for the fast neutrons sensitivity of the detector. 
\\In this paper we take into account the main materials in the Multi-Blade detector~\cite{MB_2017} listed in subsection~\ref{mbdescr}.
We calculate the probability of interaction in a medium, P(x), and the probability of deposition in the gas, P(r), as follows:

\begin{equation}
P(x) = 1 - e^{-\Sigma x}
\label{inter}
\end{equation}

\begin{equation}
P(r) = 1 - e^{-\Sigma r}
\label{escape}
\end{equation}

where x is the path length and r is the range of the particle travelling in the medium and $\Sigma = N \sigma$ is the macroscopic cross section. $\Sigma$ defines the probability per unit path length for a specific process, described by the microscopic cross section $\sigma$; $N$ is the number of nuclei per unit volume. The number of interaction is described as $\frac{I}{I_{0}} = e^{-\Sigma x}$. P(r) = P(x) for the gas.

We calculate the macroscopic cross section from the number density $n_p$ and the microscopic cross section. We used $n_p = \frac{P}{K_B T}$ for the gas and $n_p = \frac{N_A}{M} \cdot \rho$ for the solids. 
\\ In a proportional chamber the gas is the detection medium, for the Multi-Blade detector, this is an 80/20 mixture of Ar/CO$_2$ operating at atmospheric pressure and room temperature. We consider the weight $w$ and the fraction $f$ of the two compounds. The number density for the Ar/CO$_2$ is shown in table~\ref{gasnp}, while for pure materials the $n_p$ is listed in table~\ref{sol}.

\begin{table}[htbp]
\centering
\caption{\label{gasnp} \footnotesize Atomic density value for the gas Ar/CO$_2$ (80/20) in the detector.}
\smallskip
\begin{tabular}{|c|c|c|c|}
\hline
  & f & w & $n_p$ ($10^{22}/cm^3$) \\
\hline
Ar & 1 &  0.8 & 0.0212 \\
\hline
C & 1/3 & 0.2 & 0.00177 \\
\hline
O$_2$ & 2/3 & 0.2 &  0.00354\\ 
\hline
\end{tabular}
\end{table}

\begin{table}[htbp]
\centering
\caption{\label{sol} \footnotesize Atomic density value for solid pure materials in the detector.}
\smallskip
\begin{tabular}{|c|c|c|c|}
\hline
 &  $\rho $&  $M $& $n_p $ ($10^{22}/cm^3$)\\
\hline
Al  & 2.7 &  26.98 & 6.026 \\
\hline
Ti & 4.506 & 47.87 & 5.669 \\
\hline
Cu & 8.96 &63.55 & 8.491 \\ 
\hline
\end{tabular}
\end{table}

The Kapton ($\mathrm \rho$ = 1.42 $\mathrm g/cm^3$) is composed of several elements, the macroscopic cross section is  $\Sigma_{mix} = \sum_i N_i \sigma_i$, the $\mathrm{^{10}B_{4}C}$ layer ($\mathrm \rho$ = 2.45 $\mathrm g/cm^3$) as well. The n$_p $ calculation for both compounds is reported in table~\ref{kap}.

\begin{table}[htbp]
\centering
\caption{\label{kap} \footnotesize Atomic density value for the kapton and the $\mathrm{^{10}B_{4}C}$ elements.}
\smallskip
\begin{tabular}{|c|c|c|c|c|c|}
\hline
 & & $w$ & $M $ & $\rho $  & $n_p $ ($10^{22}/cm^3$) \\
\hline
& H & 0.0264 &  1.01 & 0.037 &  2.23 \\
\cline{2-6}
Kapton & C & 0.6911 & 12.01 & 0.981 &  4.92 \\
\cline{2-6}
& N & 0.0733 & 14.01  & 0.104 & 4.47 \\
\cline{2-6}
& O & 0.2092 &15.99 &0.297 & 1.12  \\ 
\hline
\hline
$\mathrm{^{10}B_{4}C}$ & B & 0.7826 & 10.81 & 1.972 &  10.98 \\
\cline{2-6}
& C & 0.2174 & 12.011 & 0.548 &  2.75 \\
\hline
\end{tabular}
\end{table}

\begin{figure}[htbp]
\centering
\includegraphics[width=1\textwidth,keepaspectratio]{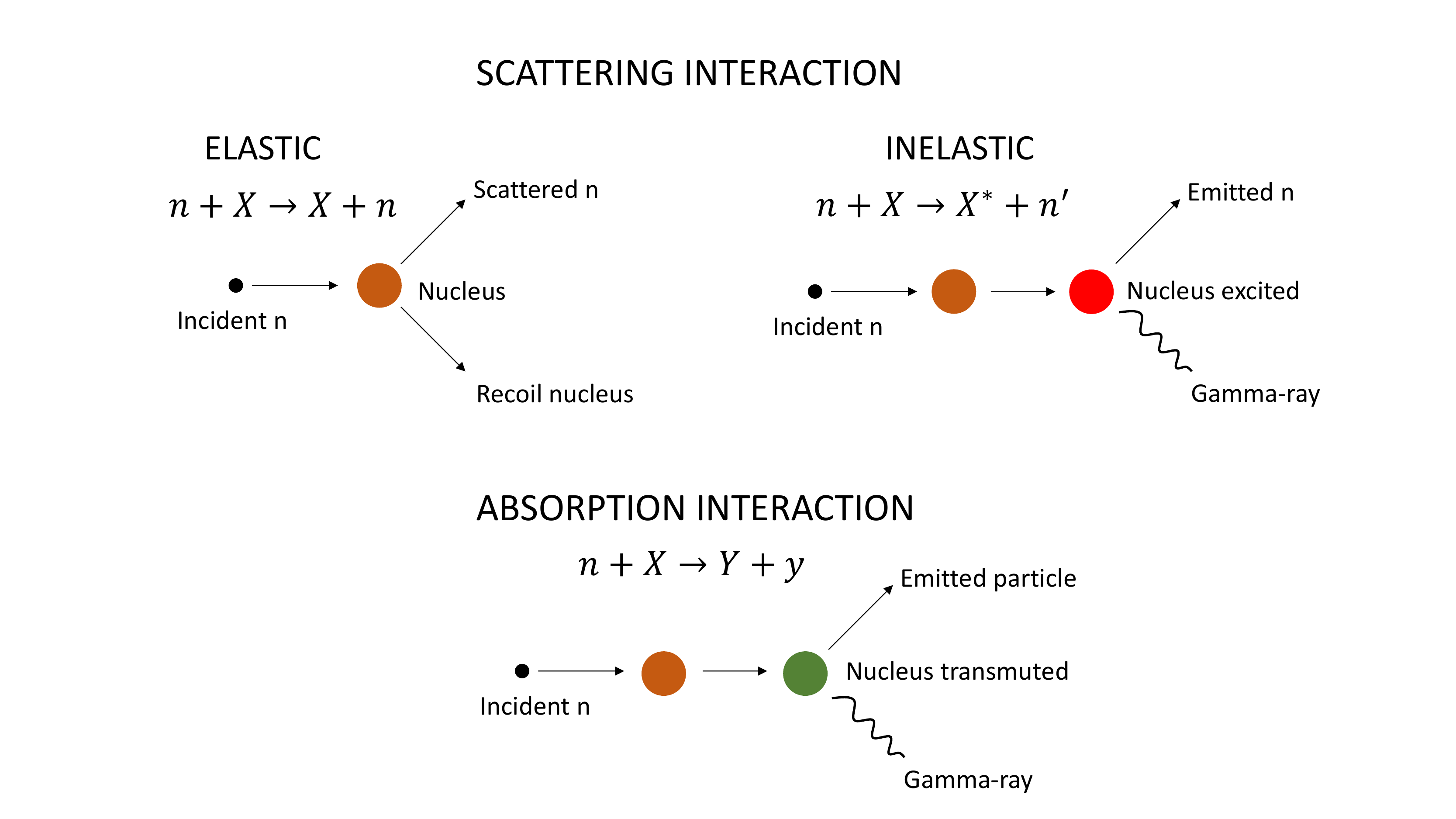}
\caption{\label{interaction} \footnotesize  A schematic view of the interactions of the fast neutron with matter. Elastic scattering interaction (top-left): the recoil nucleus and a scattered neutron is emitted. Inelastic scattering (top-right): the nucleus goes into an excited state and emits one or more gamma-rays. Absorption interaction (bottom) includes all the processes with heavy particles or gamma-rays as a yield.}
\end{figure}

A neutron can have many types of interaction with a nucleus. In the detector these can give rise to charged products that can release an amount of energy in the detection medium and can be interpreted as real thermal neutron events, instead of being rejected. For this work we consider the fast neutron energy range between 1 to 10 MeV, see subsection~\ref{subsources}. In this range several interactions between neutrons and nucleus may occur. Our focus is on two main types: \textit{Scattering} and \textit{Absorption}. 
\\ When a neutron is scattered by a nucleus, the scattering interaction transfers some portion of the neutron kinetic energy to the target nucleus resulting in a \textit{recoil nucleus}. The total kinetic energy of the neutron-nucleus system remains unchanged by the interaction, i.e. the Q-value of the reaction is zero.
%therefore the sum of the kinetic energies of the reaction products (recoil nuclues and scattered neutron) must equal that carried in by the incident neutron, see figure~\ref{interaction}. 
In the considered energy range the maximum possible recoil energy of the nucleus is 

\begin{equation}
E_R = \frac{4A}{(1+A)^2}  E_n
\label{e_recoil}
\end{equation}

where A is the mass of the target nucleus , $E_R$ the recoil nucleus kinetic energy and $E_n$ the incoming neutron kinetic energy~\cite{DET_knoll}. 
\\ Referring to figure~\ref{interaction}, the scattering interaction also includes the \textit{inelastic scattering} in which the recoil nucleus goes to one of its excited states during the collision and then de-excites emitting one or more gamma rays or internal conversion electrons. 
\\ If the neutron is absorbed or captured, instead of being scattered, a variety of processes may occur. The nucleus may rearrange its internal structure and release gamma rays (n,$\gamma$) or charged particles with the more common ones being protons (n,p), alpha particles (n,$\alpha$) and deuterons. For this energy range only p and $\alpha$ are relevant. We include all these kind of interactions in absorption processes~\cite{n_interaction_matter}. A sketch of these reactions is shown in figure~\ref{interaction}.

\begin{figure}[htbp]
\centering
\includegraphics[width=1\textwidth,keepaspectratio]{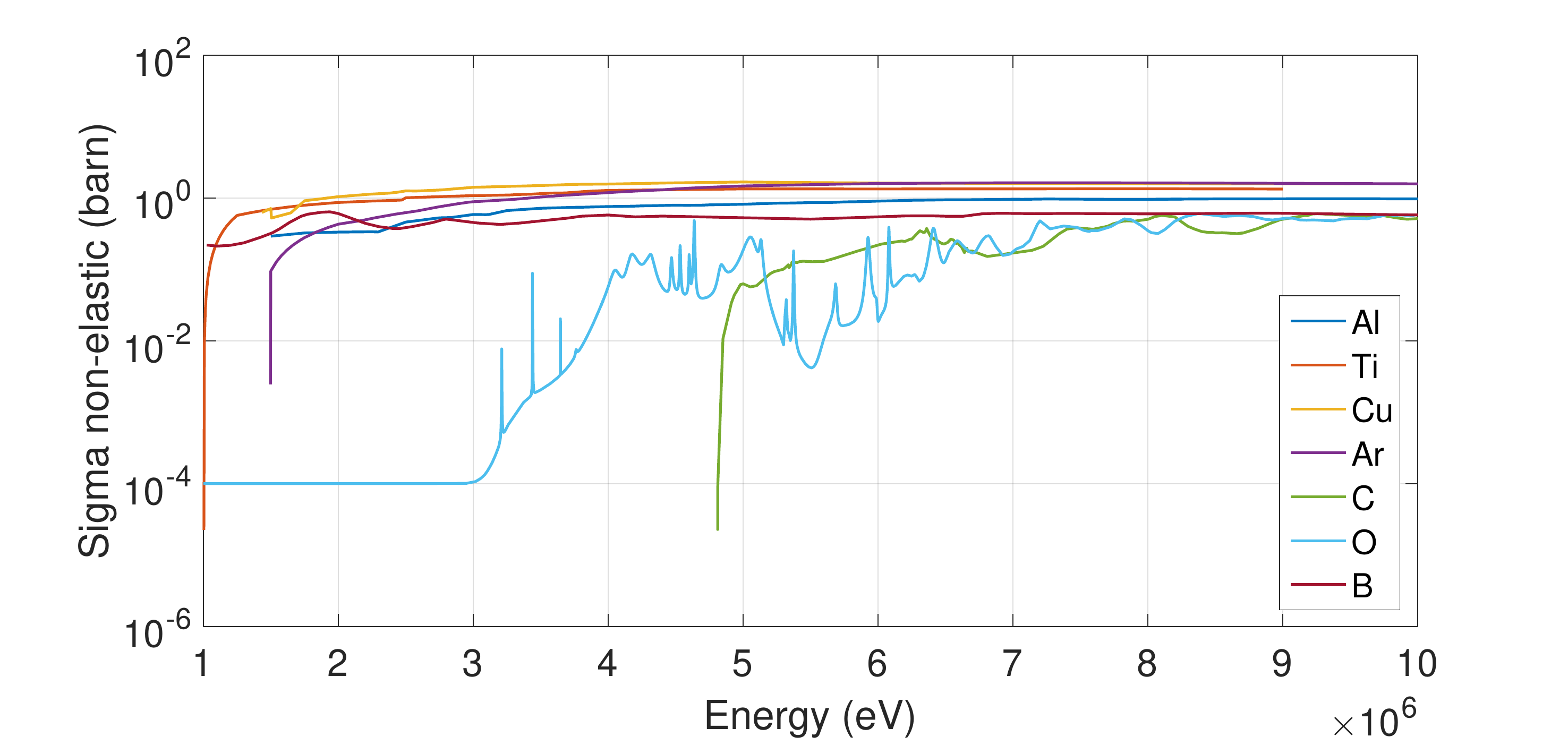}
\includegraphics[width=1\textwidth,keepaspectratio]{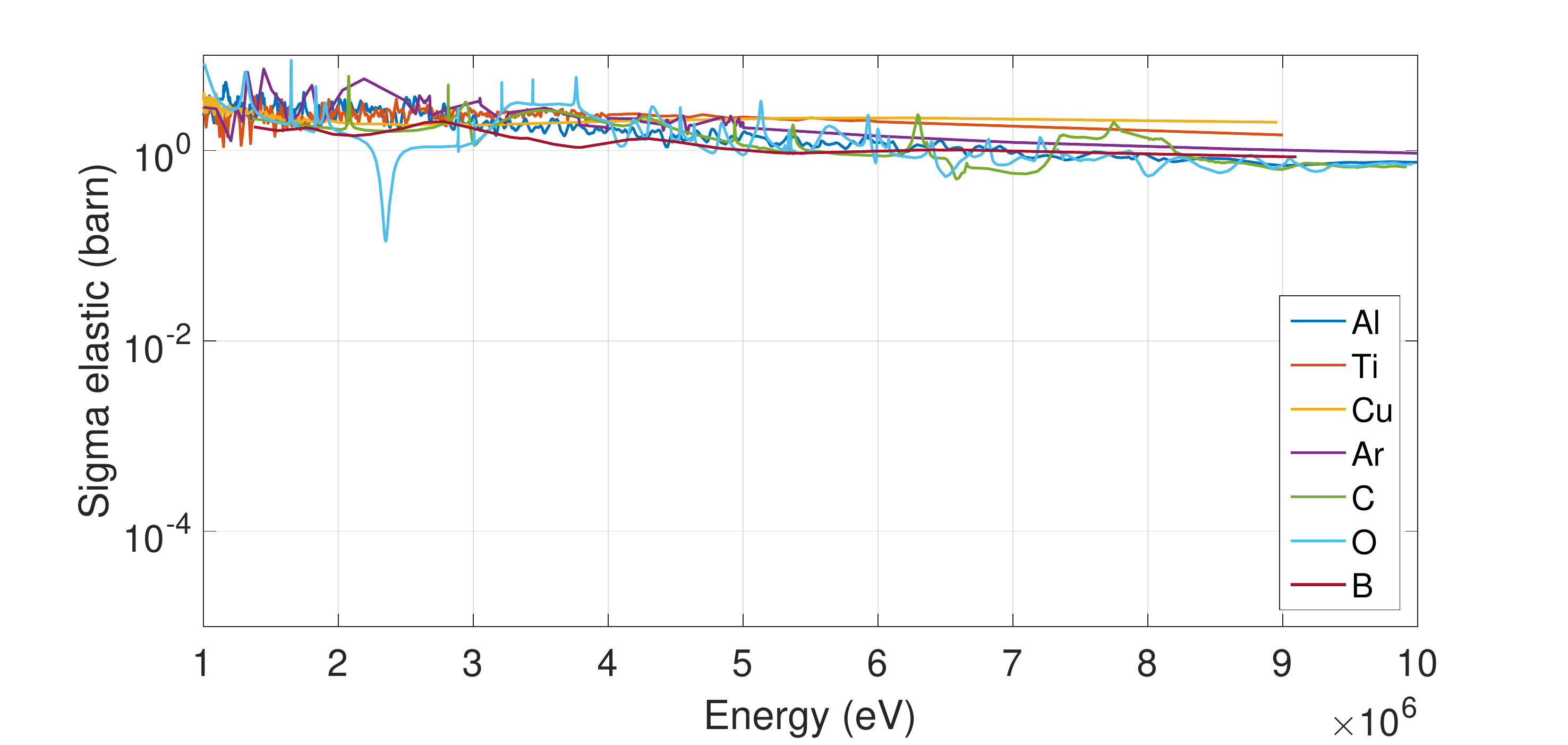}
\caption{\label{snoel} \footnotesize  Non-elastic (top) and elastic (bottom) cross sections for the materials of the Multi-Blade detector. The considered energy range is 1-10 MeV, matching the emission spectrum of the sources used in the measurements.}
\end{figure}

An important difference between the fast neutron interaction and the $\gamma$-rays interaction is the final products. When a $\gamma$-ray interacts with a medium, it can generate a detectable signal, via an electron, typically in a photoelectric or Compton interaction, while for fast neutrons protons, $\alpha$ and heavier particles can also be emitted.
\\The range of protons and heavier particles in a medium is typically much shorter than that of electrons for a given energy. Therefore the amount of energy released per unit path in the gas is much larger for heavy particles than for electrons \cite{MG_gamma}. 
\\ The fast neutron interaction can give rise to $\gamma$-rays as well which undergo the process as described above. In $\mathrm{^{10}B}$-based detector, it has been denoted that the detection probability of $\gamma$-rays is low of the order of $10^{-8}$, as reported in~\cite{MG_gamma, MB_2017}. Therefore a larger contribution to the sensitivity to fast neutron is mainly expected from (n, particles) reactions and the elastic scattering. The contribution from the inelastic scattering and (n,$\gamma$) interactions is negligible. 
\\ We define \textit{Non-Elastic} interaction as the sum of inelastic and absorption processes. Figure~\ref{snoel} shows several non-elastic (top) and elastic (bottom) cross sections, in the energy range 1-10 MeV~\cite{NIST}.

In the considered energy range, the elastic scattering interaction has the largest cross sections and the higher probability of deposition in the gas P(r) (eq.~\ref{escape}). The inelastic scattering and (n,$\gamma$) processes are comparable for each considered material and they have the lowest probability. 
\\The most relevant absorption processes we take into account are~\cite{Al_np, Ti_np, O_nalpha}: 

\begin{equation}\label{al_p}
^{27}Al(n,p)^{27}Mg \quad Q = -1.8MeV
\end{equation}

\begin{equation}\label{ti_p}
^{48}Ti(n,p)^{48}Sc \quad Q = -3.19 MeV
\end{equation}

\begin{equation}\label{o_alpha}
^{16}O(n,\alpha)^{13}C \qquad Q = -2.2 MeV
\end{equation}

\begin{figure}[htbp]
\centering
\includegraphics[width=1\textwidth,keepaspectratio]{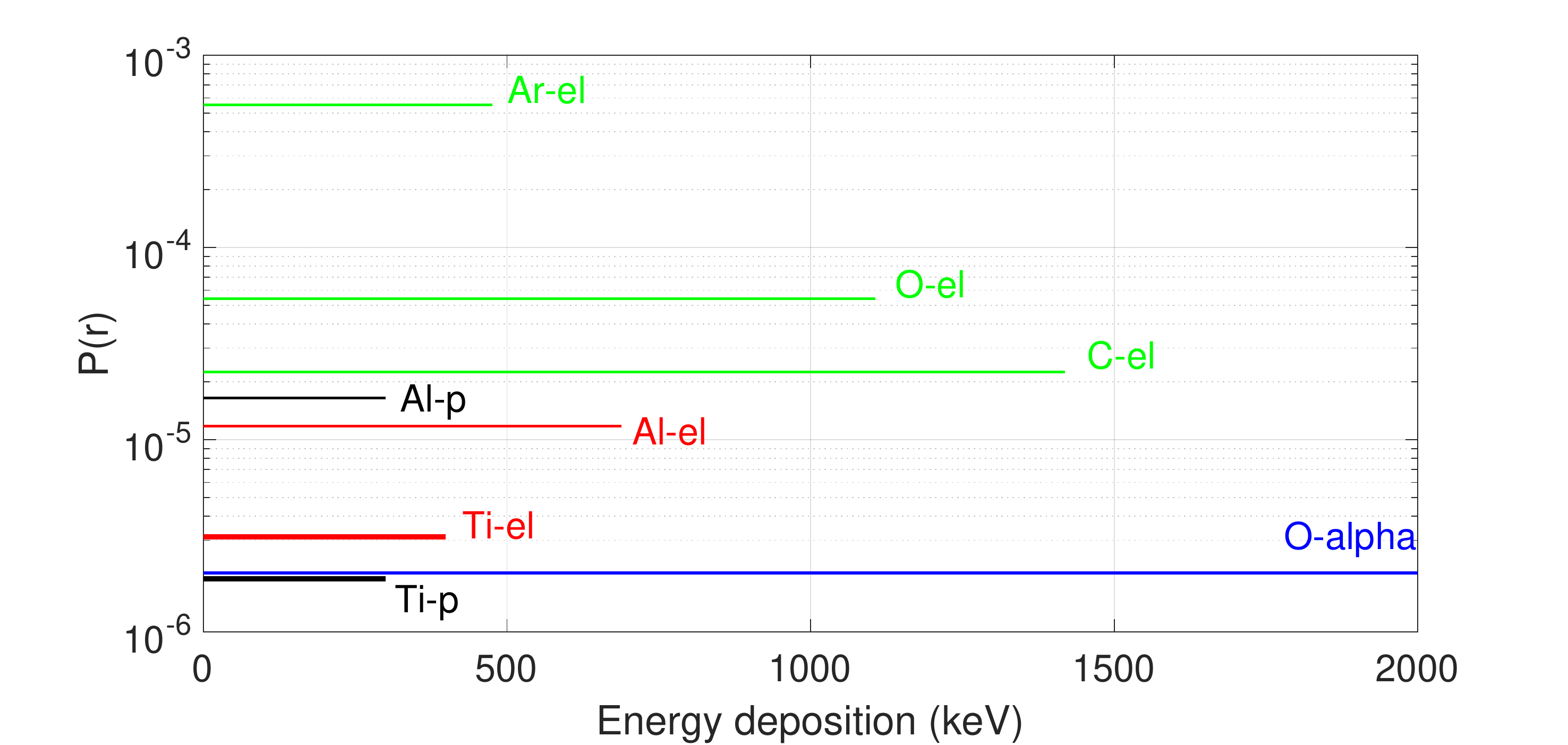}
\caption{\label{theo_el} \footnotesize Theoretical calculation of the maximum energy deposition in the gas and its probability for the different components of the detector. The contribution of the gas is dominant with respect to that of solids.}
\end{figure}

In general these absorption processes are: $$n + X \rightarrow Y + y$$ For each reaction we calculate the energy threshold ($\mathrm E^{th}_n$) for the incoming neutron to generate the reaction as 

\begin{equation}
E^{th}_n = - \Big(1+ \frac{m_n}{M_X} \Big)\cdot Q
\label{e_th}
\end{equation}

and the energy of the produced particles ($\mathrm E_y$) from the definition of the Q-value and the conservation of energy and momentum as 

\begin{equation}
E_y =  \Big( \frac{M_Y}{M_Y + m_y}\Big)\cdot Q_n
\label{e_y}
\end{equation}

where $Q_n = E_n +Q$ is the sum of the incoming energy of neutrons and the Q-value.
\\ In order to compare our results to a theoretical prediction, we chose to an incoming neutron energy of $\mathrm E_n = 5$ MeV for our calculations, because it is the average value of the energy range of interest and is approximately the average of the emission spectrum of the fast neutron sources used for the measurements (see section~\ref{subsources}).
We calculate the probability of deposition P(r) (eq~\ref{escape}) summarized in table~\ref{tabsp}, considering the average of the cross sections $\sigma$ for each interaction in the given energy range and the number density n$_p$ for each material, in the detector. 
\\ For the elastic scattering and the absorption interactions we calculate $\mathrm E_R$ and $\mathrm E_y$ respectively, as described above. We used these energies as input in a SRIM simulation~\cite{ MISC_SRIM2010, MISC_SRIM1998} estimating the maximum range, r. We consider, instead, the nominal thickness x of the medium for the inelastic scattering interactions, because the yields of this process are $\gamma$-rays. 
\\ The values are shown in table~\ref{tabsp} and the individual contribution of the elements for the elastic and the absorption interactions are shown in figure~\ref{theo_el}.

\begin{table}[htbp]
\centering
\caption{\label{tabsp} \footnotesize Macroscopic cross sections and probability of deposition in the gas values for the maximum range, r, and the thickness, x, given by the detector geometry.}
\smallskip
\begin{tabular}{|c|c|c|c|}
\hline
\multicolumn{4}{|c|}{\textbf{Elastic Interaction}}\\
\hline
 &  $\Sigma (10^{-3}/cm)$ & $r (cm)$ & $P(r)(\%)$  \\
\hline
 Al & 139.2 & 0.0000874  & 0.00117 \\
 \hline
Ti & 134.9 & 0.0000232 &  0.000313\\
\hline
Cu & 224.2 &  0.0000089 &  0.000199 \\
\hline
$\mathrm{Ar/CO_2} $ & 0.79 & 0.803 & 0.0629 \\ 
\hline
$\mathrm{^{10}B_{4}C}$ & 202.2 & 0.000148 & 0.0029\\
\hline\hline
\multicolumn{4}{|c|}{\textbf{Inelastic Interaction}}\\
\hline
 &  $\Sigma (10^{-3}/cm)$ & $x (cm)$ & $P(x)(\%)$  \\
\hline
 Al & 41.83 & 2.295  & 9.48 \\
 \hline
Ti & 44.67 & 2.295 &  10.04\\
\hline
Cu & 94.14 &  0.021 &  0.197 \\
\hline
Kapton & 8.259 &  0.046 &  0.038 \\
\hline
$\mathrm{Ar/CO_2} $ & 0.188 & 8.032 & 0.0168 \\ 
\hline
$\mathrm{^{10}B_{4}C}$ & 58.95 & 0.008 & 0.046\\
\hline\hline
\multicolumn{4}{|c|}{\textbf{Absorption Interaction}}\\
\hline
 &  $\Sigma (10^{-3}/cm)$ & $r (cm)$ & $P(r)(\%)$  \\
\hline
 Al(n,p) & 44.09 & 0.0081  & 0.0016 \\
\hline
Ti(n,p)& 44.67 & 0.0026 &  0.00018\\
\hline
O(n,$\alpha$) & 103.1 &  0.803 &  0.0002 \\
\hline
\end{tabular}
\end{table}

A direct comparison between the materials is possible because the probability of deposition in the gas is normalized by the density. An interaction in the solid generates charged particles that have to escape the surface and reach the gas to be detected. With the SRIM calculation we quantify the portion of surface involved in the process (maximum range r). A direct comparison is then possible between solids and gas components as well.
\\ Referring to figure~\ref{theo_el}, the energy released by the elastic process of the gas elements is dominant ($\approx$ 40 times larger) compared to the same contribution of the several solid components. The energy deposited by the various elastic reactions is 1-2 orders of magnitude larger than the contribution of the absorption interactions.
\\ Although the $\mathrm{^{10}B_{4}C}$ layer is not present in the detector or this experimental setup, from the calculation we learn that its contribution would be approximately 1/20 of that of the $\mathrm{Ar/CO_2} $.
\\ The comparison between Aluminium and Titanium is crucial for the Multi-Blade design, because it concerns the choice of the blade material, aiming at an improved detector performance~\cite{MB_2017}. A difference emerges from the calculation of the probability of deposition for elastic scattering. For Aluminium, P(r) is approximately 4 times larger than for Titanium and about 6 times higher than for Copper. Cu has one of the largest microscopic cross sections among the studied materials (subsection~\ref{mbdescr}) in the energy range (subsection~\ref{subsources}), but it is the heaviest element among them. It is then less probable for a particle to escape the surface and release energy in the gas. Referring to the absorption processes described above, the deposition probability of the (n,p) process for Aluminium is 1 order of magnitude larger than for Titanium. 

\begin{figure}[htbp]
\centering
\includegraphics[width=1\textwidth,keepaspectratio]{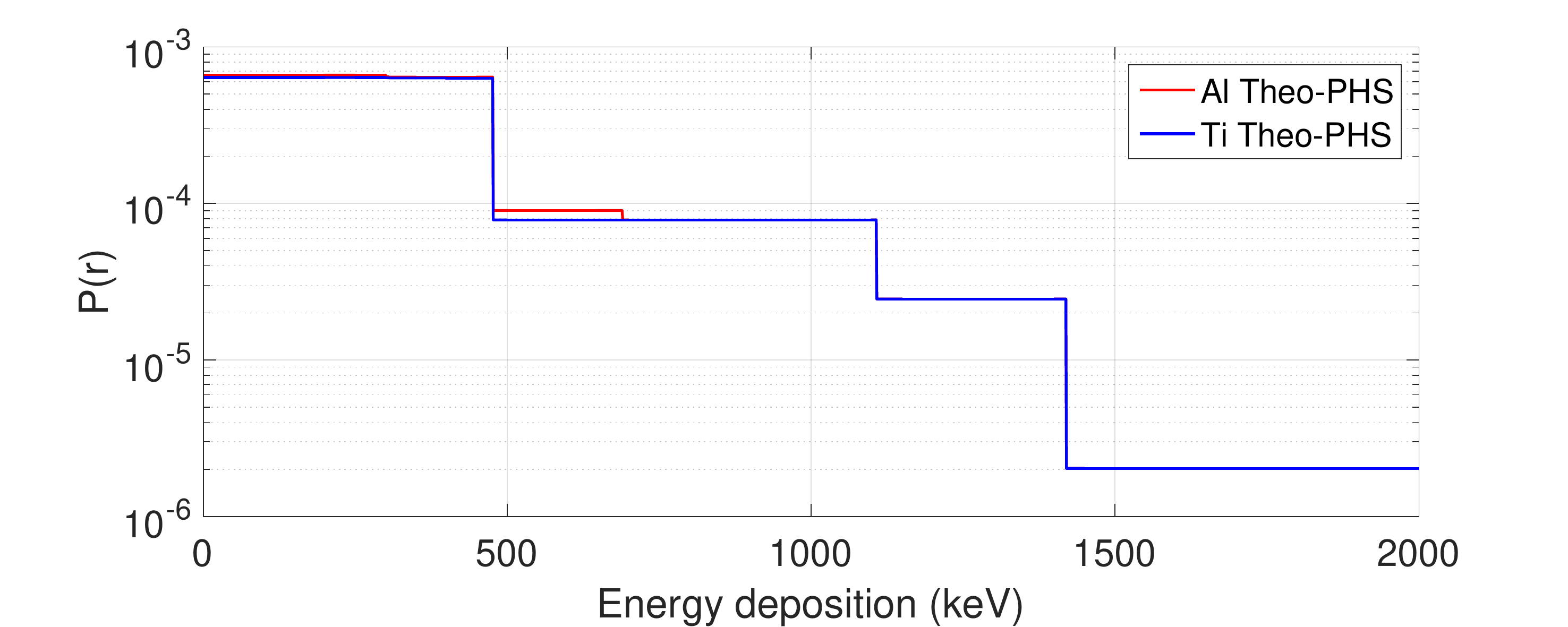}
\caption{\label{theo_phs} \footnotesize Theoretical representation of the energy deposition in gas in case of Aluminium and Titanium blades assuming the same probability, P(r), in all the possible energy range defined by SRIM calculation.}
\end{figure}

Figure~\ref{theo_phs} depicts the sum of the individual energy deposition in the gas according to figure~\ref{theo_el} distinguishing between Aluminium and Titanium, as described in subsection~\ref{subsetup}. The plot is a theoretical representation of a Pulse Height Spectrum (PHS) assuming that a particle has the same probability, P(r), to deposit any energy, in all the possible energy range defined by SRIM calculation.

\begin{figure}[htbp]
\centering
\includegraphics[width=0.8\textwidth,keepaspectratio]{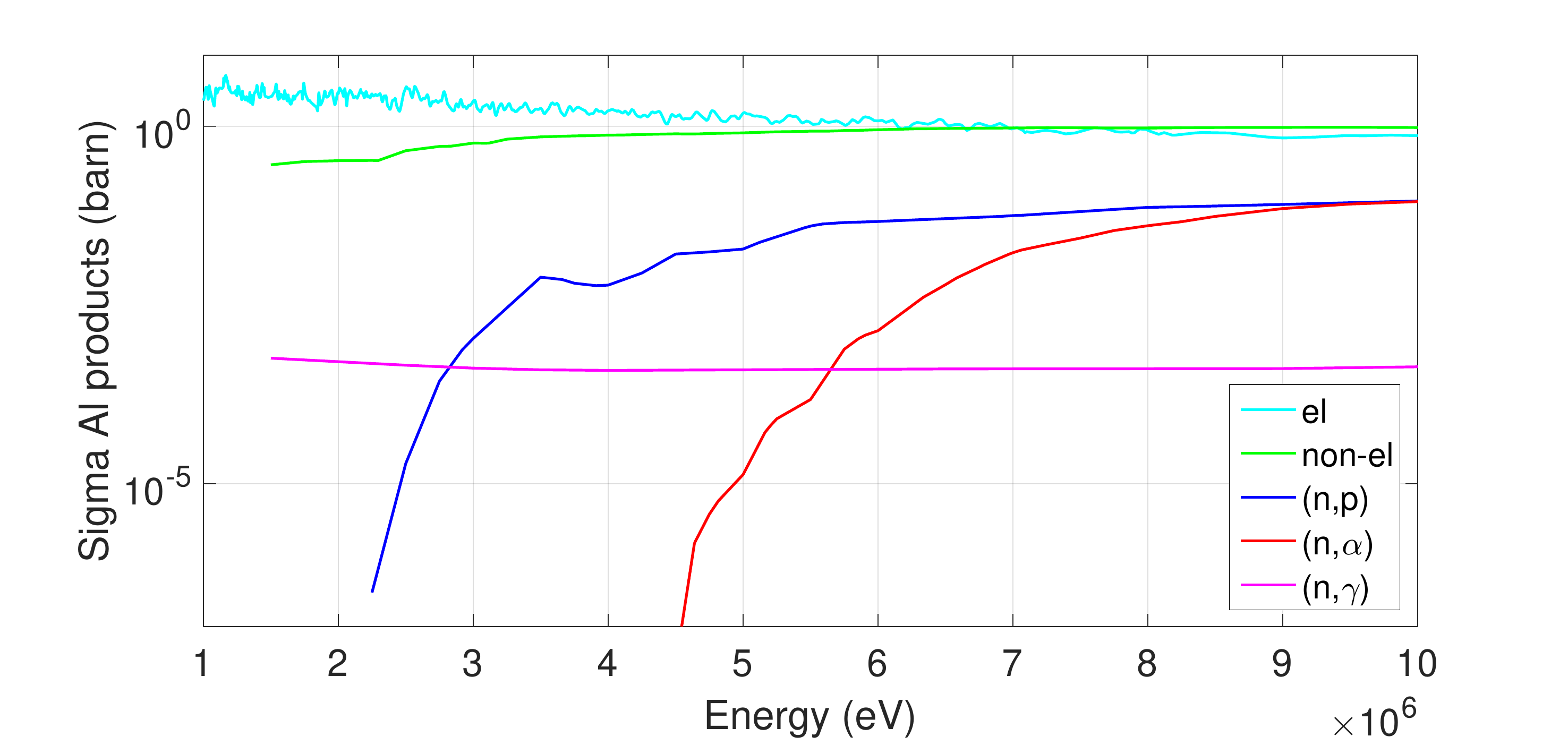}
\includegraphics[width=0.8\textwidth,keepaspectratio]{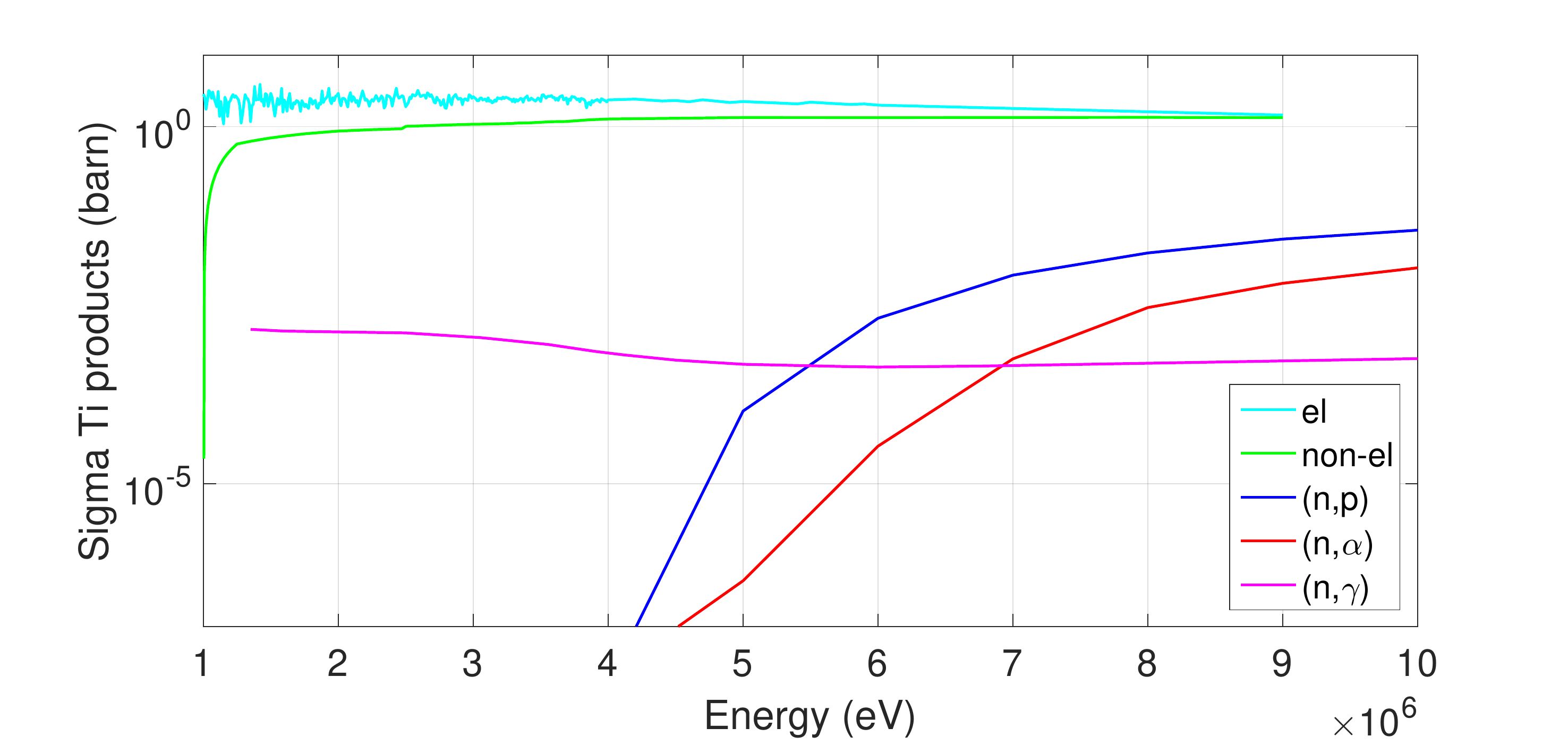}
\caption{\label{sigmaAlTi} \footnotesize Comparison between elastic and total non-elastic cross sections. The components of the non-elastic cross section are also shown:  $(n,p)$, $(n,\alpha)$, $(n,\gamma)$ processes both for Aluminium (top) and Titanium (bottom). In Aluminium the $(n,p)$ process occurs at 3 MeV, while in Titanium it has the same probability at 6 MeV.}
\end{figure} 

We compare the cross sections for the three most relevant absorption processes: $(n,p)$, $(n,\alpha)$, $(n,\gamma)$ and the total non-elastic cross section of two materials, figure \ref{sigmaAlTi}.
\\ The $(n,p)$ process in Al occurs approximately at $\approx$3 MeV with a probability  $\sim10^{-3}$, while for Ti the cross section of this process has the same order of magnitude at $\approx$6 MeV. Note that the neutron sources we used for the measurements have an intensity of the emission spectrum which falls off at larger energies up to 10 MeV.

\section{Geant4 simulations}\label{simu}

\begin{figure}[htbp]
\centering
\includegraphics[width=0.6\textwidth,keepaspectratio]{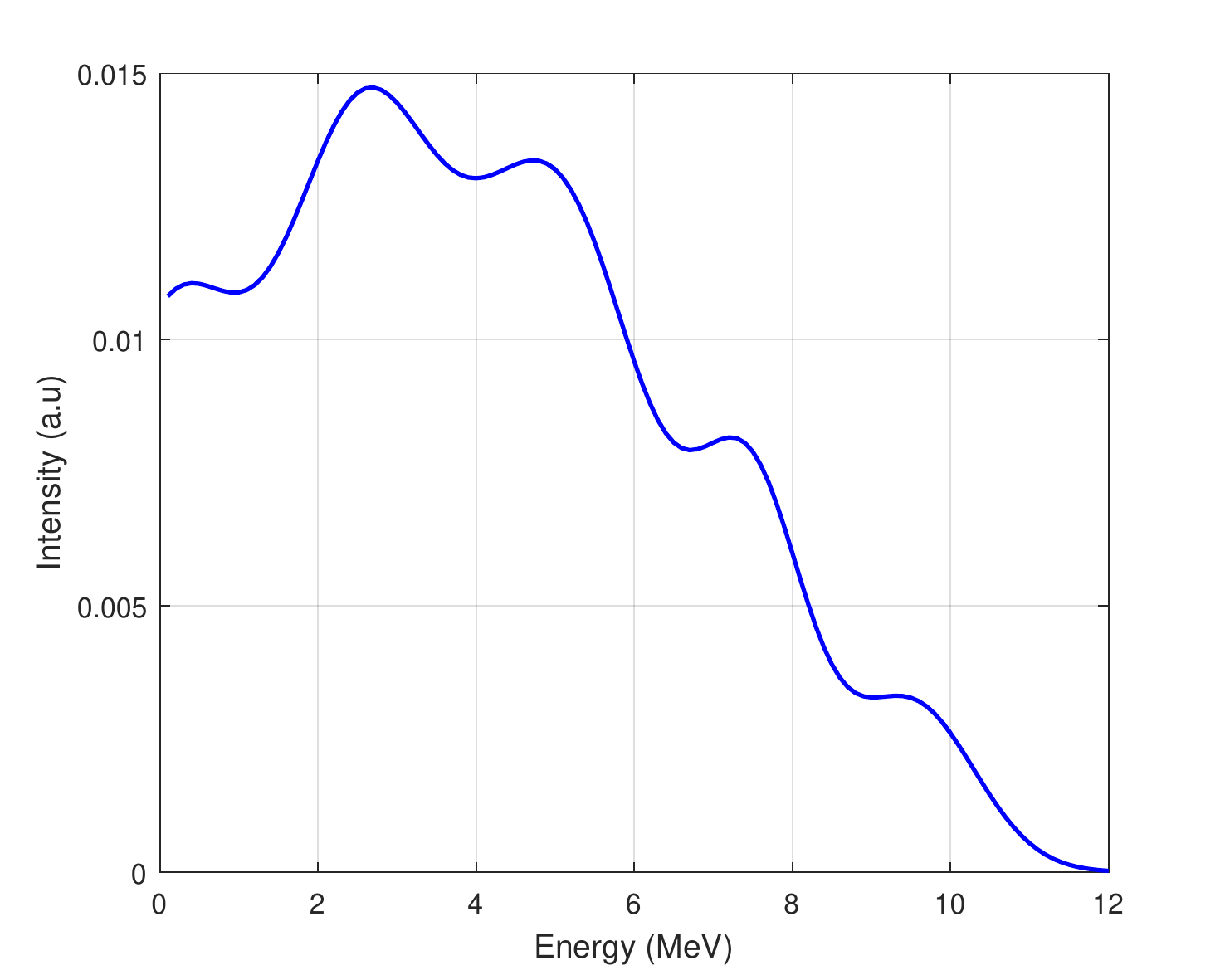}
\caption{\label{pubegauss} \footnotesize $^{238}$Pu/Be-like fast neutron energy spectrum
  used as input for the Geant4 simulations.}
\end{figure}

As a qualitative cross validation of the theoretical conclusions on the shape
of the PHS, a Geant4~\cite{g1,g2,g3} simulation is
performed with a ``realistic'' energy spectrum of the neutron
source. In particular, an approximate neutron energy spectrum of a $^{238}$Pu/Be
source is used as input (see figure~\ref{pubegauss}). It is modelled using several Gaussian distributions, referring to the work on the characterization of the $^{238}$Pu/Be and $^{241}$Am/Be source~\cite{PuBe_AmBe}.

\begin{figure}[htbp]
\centering
\includegraphics[width=0.5\textwidth,keepaspectratio]{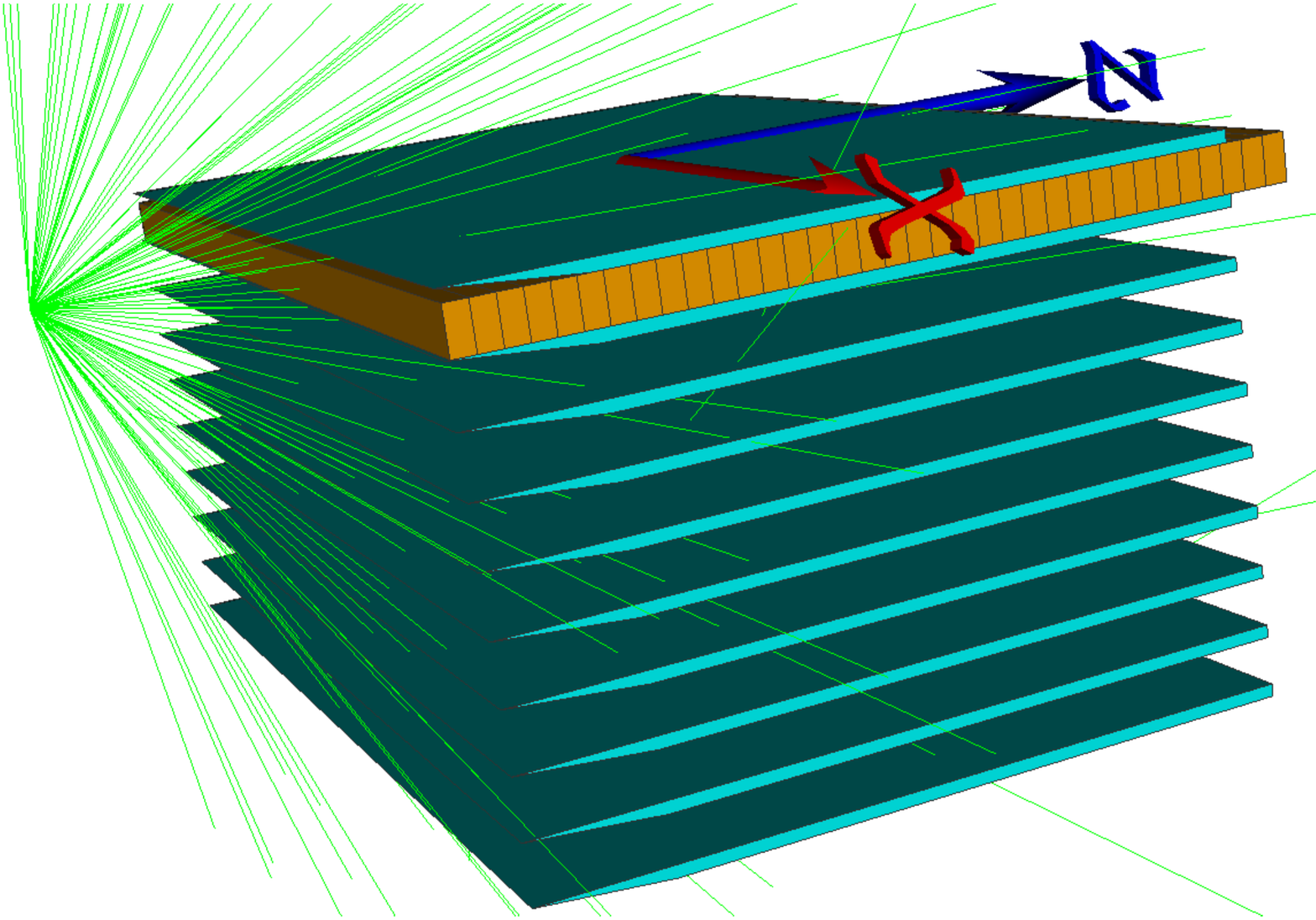}
\caption{\label{geoMB} \footnotesize Approximated Multi-Blade geometry
  with source placement as in the experimental setup. The orange volumes represent the wire
  voxels. In green are the tracks of the primary neutrons generated at the
  source.}
\end{figure}

The implemented detector geometry and the source placement match the experimental setup and are
depicted in figure~\ref{geoMB}. The detector consists of ten blades
arranged in parallel, with a segmented gas region between them. The
segmentation resembles the
sensitive volumes around the wires and readout of the experimental conditions. The model
is an approximation of the actual detector in the sense that the
blades are parallel instead of having the fan-like arrangement used
for thermal neutron detection, a compromise
that facilitates the implementation of the gas volume segmentation without an impact on the
fast neutron detector response. 

The neutron source emits isotropically and is placed 5~cm from the
edge of the readout wire plane as shown in figure~\ref{geoMB}. The
particle generator produces only neutrons, without photons or other
sources of background included. No detector vessel, kapton, tungsten wires or copper strips are present in the particular geometry
model either. Their impact has been studied in a separate set of
Geant4 simulations and as the results have fully matched the
experimental findings, a simplified detector geometry model is
utilised for the results presented here.

The material chosen for the gas composition is a mixture of Ar and
CO$_2$ with an 80:20 by mass ratio. The material of the blades is
either pure Al or Ti of natural isotopic composition with a
poly-crystalline structure. The latter is enabled via the use of
the NCrystal library~\cite{ncrystal} and will properly treat the interactions
of low energy neutrons with the crystalline material of the blades, in
case thermalization occurs in the setup. In terms of physics
list, the standard QGPS\_BIC\_HP covers the processes of interest.

Assuming perfect energy resolution, the energy deposited by any particle in the gas volume is collected by
applying the experimental condition of a 100~keV threshold per wire
volume. All wire volumes are included in the formation of the
simulated PHS to gain a statistical advantage, as the gas
amplification stage is excluded, thus equalizing their response. 

\subsection{Individual contributions to the PHS}\label{sim_contribute}

\begin{figure}[htbp]
\centering
\includegraphics[width=1\textwidth,keepaspectratio]{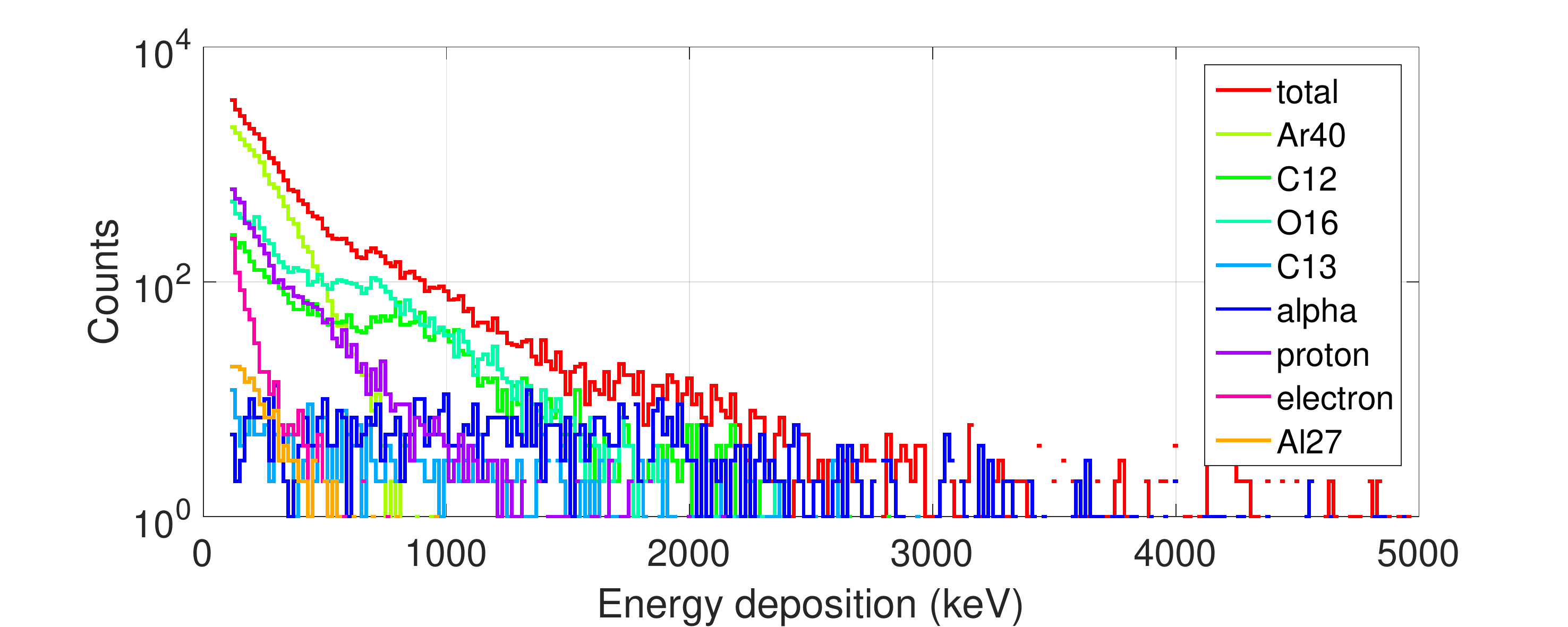}
\includegraphics[width=1\textwidth,keepaspectratio]{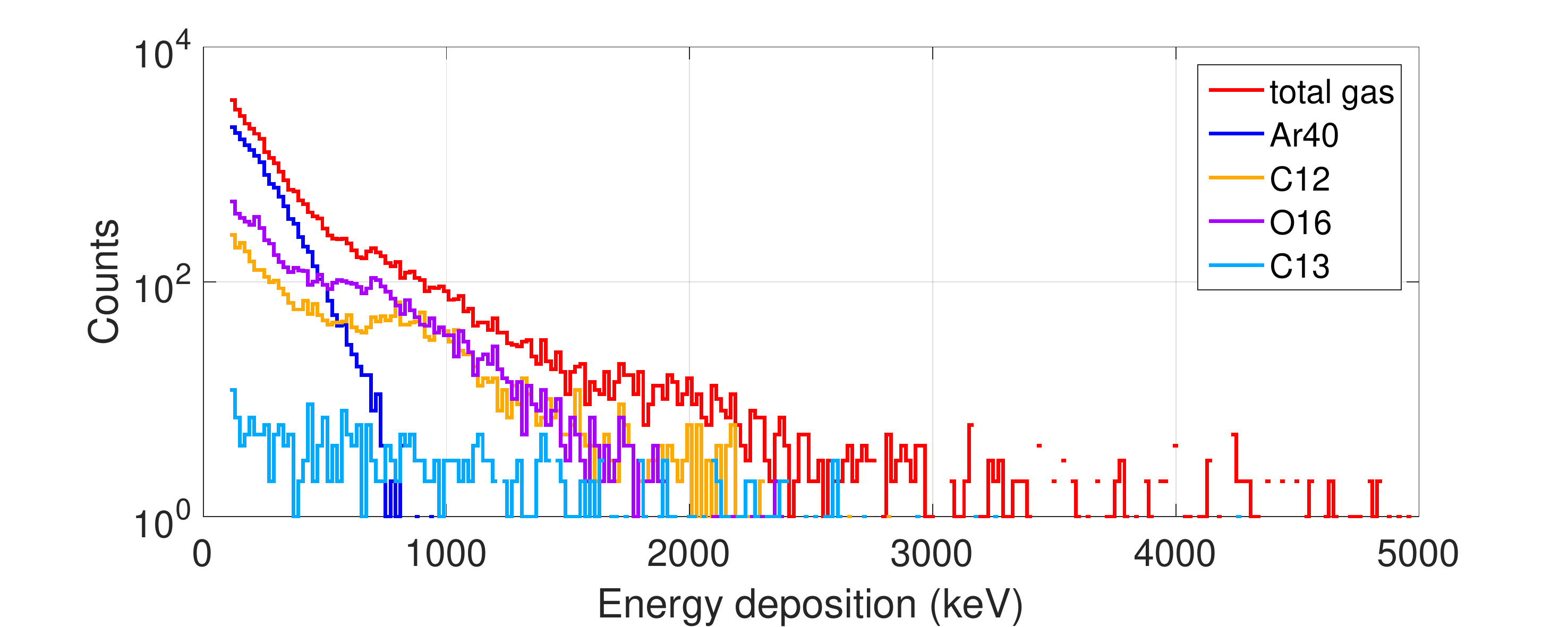}
\caption{\label{sim_gas_edep} \footnotesize Simulated energy
  deposition of the main detector components (top) and the gas components (bottom).}
\end{figure}

The individual contributions to the deposited energy in the gas are
identified and the shape of the PHS can be explained by them. Figure~\ref{sim_gas_edep} demonstrates in a quantitative way that the
energy depositions from $^{12}$C and $^{16}$O are the reason for the
spectrum bump around 800~keV. The $^{40}$Ar contribution drops
relatively sharply without structure until 1000~keV, while the
$^{13}$C spectrum is flatter and extends up to 3000~keV. The gas
contributions dominate the spectrum until 1000~keV, while other
particles, like protons and $\alpha$ are responsible for the shape
of the spectrum until 5000~keV. 

The elastic interactions contribute to the
energy deposition spectrum in a wide energy region, while all inelastic processes can be attributed to the electron energy deposition in the gas, whose contribution does not extend above 500 keV.
These findings are consistent with and validate the
prior theoretical analysis of cross sections in section~\ref{theocalc}. 
The contributions from the blade atoms are minimal and sharply drop in
the low values of the spectrum, as seen in figure~\ref{sim_gas_edep}. 

\begin{figure}[htbp]\centering
\includegraphics[width=1\textwidth,keepaspectratio]{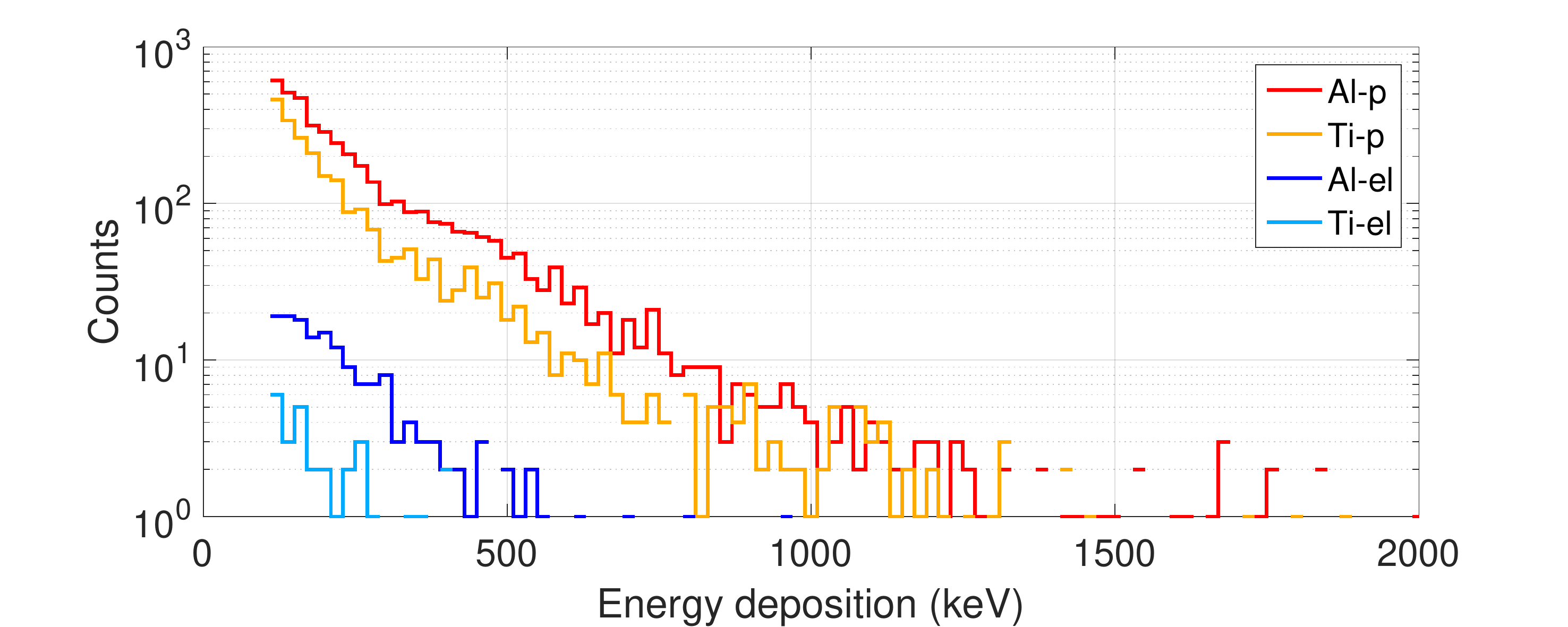}
\caption{\label{proton} \footnotesize Proton energy loss in the gas and nucleus recoil energy from elastic scattering for Aluminium and Titanium blade material.}
\end{figure}

A closer study at the individual components between the spectra for Al and Ti blades leads to the conclusion that the difference is mostly attributed to
the proton energy loss in the gas and the elastic reaction (see figure~\ref{proton}), in agreement with the analysis of section~\ref{theocalc}. 
Moreover for the same number of initial events, the ratio between the PHS obtained with Al and Ti is (2$\pm 1$), in agreement both with the calculation, section~\ref{sp}, and with the measurements shown in section~\ref{materiali}.  

\section{Measurements}\label{measure}

The measurements have been performed at the Source Testing Facility (STF)~\cite{IAEA-STF} at Lund University in Lund (SE).

\subsection{Verification of lack of thermal neutron sensitivity}

\begin{figure}[htbp]
\centering
\includegraphics[width=1\textwidth,keepaspectratio]{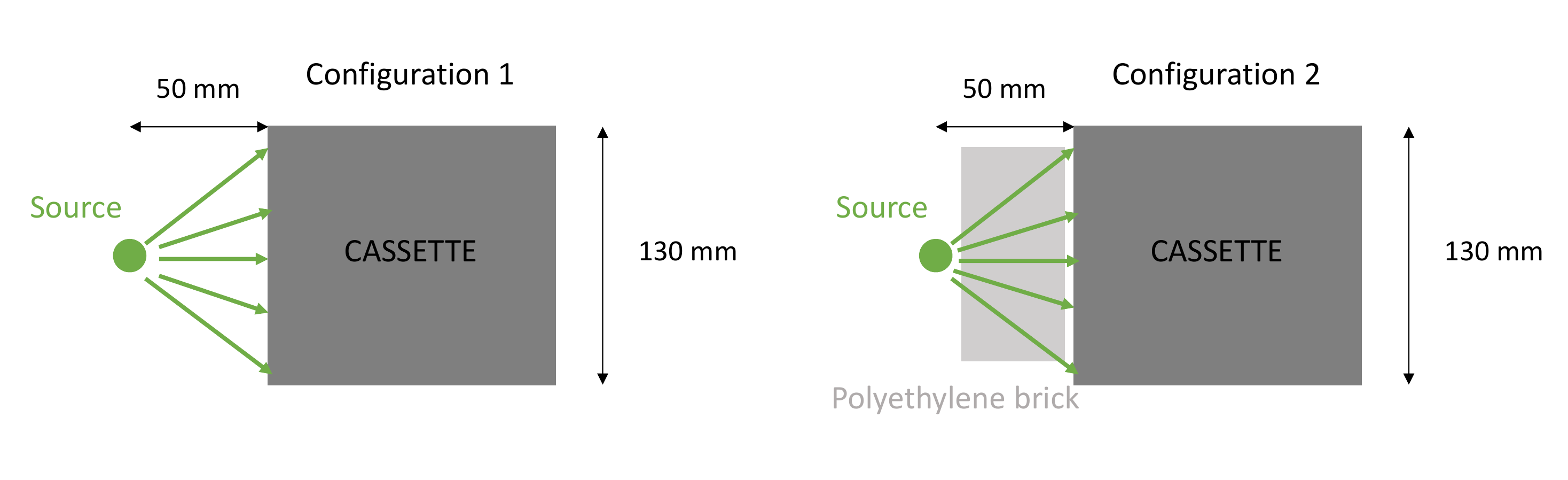}
\caption{\label{setup_brick} \footnotesize Sketch of the experimental setup with (right) and without (left) polyethylene brick in between.}
\end{figure}

Before proceeding with fast neutron measurements, a study of the thermal neutron contribution was performed to ensure that it does not affect the measurements. We compare the measurements with and without a polyethylene brick between the source and the detector in order to compare thermalized and fast neutron flux. A sketch of these configurations is shown in figure~\ref{setup_brick}, while all other measurements were performed in the Configuration 1.

\begin{figure}[htbp]
\centering
\includegraphics[width=1\textwidth,keepaspectratio]{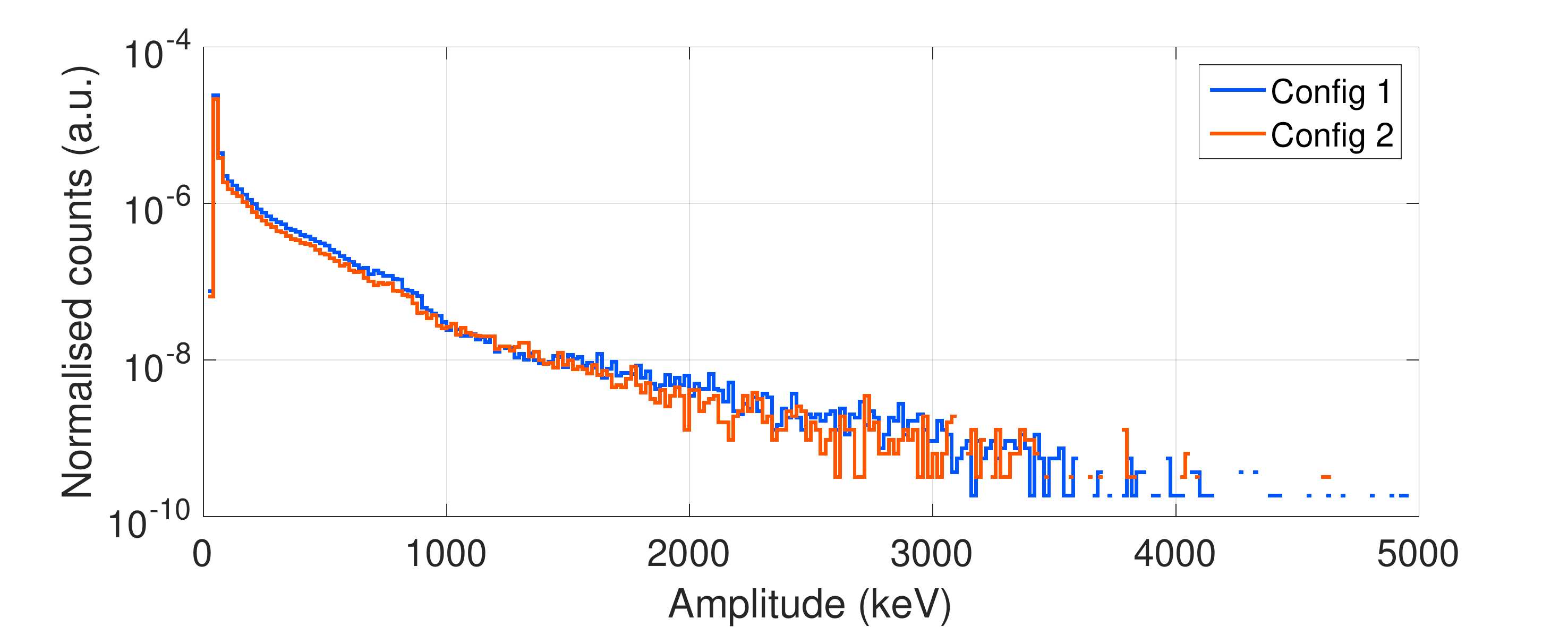}
\caption{\label{config} \footnotesize PHS acquired in the configuration 1 (blue curve) and with a polyethylene brick between the source and the detector, configuration 2 (red curve).}
\end{figure}

A PHS for the two configurations was recorded (see figure~\ref{config}). For these measurements we used the $^{238}$Pu/Be source. To get the PHS we sum all the counts recorded with the four wires in the middle of the cassette in order to increase the statistics and to take into account only wires with approximately the same gas gain. Due to the geometry, the detector has a variable gain across the wire plane as described in~\cite{MB_2017}. 
\\ In order to calibrate the measured spectra, a measurement with another cassette with the Boron layer was performed in order to measure the alpha peak of the 94\% branching ratio of the neutron capture reaction, corresponding to an energy of 1470 keV and the center of the peak is used to convert the PHS X-axis from ADC levels to energy. This calibration method is applied in all the following plots.
\\ In order to understand if we are sensitive to the thermal neutron contribution with no $\mathrm{^{10}B_{4}C}$ layer on the blade, we compare the PHS obtained in the two considered cases in figure~\ref{setup_brick}. Fewer neutrons reach the detector in the Configuration 2 (figure~\ref{setup_brick}) as they are scattered and thermalized by the polyethylene brick. We apply an arbitrary scaling factor of $\approx$2 to the PHS obtained in this case to compare with the PHS acquired in the Configuration 1 (see figure~\ref{config}). No difference emerges from this comparison, proving that we are not sensitive to thermal neutrons. As expected by removing the $^{10}$B layer no other thermal contributions are involved in the measurements.

A further test to identify the thermal and the fast neutron signal signature is the multiplicity. When a neutron is converted into charged particles, they travel in the gas releasing energy on their path. This energy can be collected on wires and strips. The number of wires or strips involved in the detection of a single event defines the multiplicity of such event. When a neutron is converted in the $^{10}$B$_4$C layer the event is recorded in $99.5 \%$ of cases on no more than 2 wires and 3 strips as described in~\cite{MB_2017}. Multiplicity for fast neutrons and gammas is in general higher, a characteristic that can be used to discriminate against background. A summary of multiplicity values is depicted in table~\ref{multiplicity}.

\begin{table}[htbp]
\centering
\caption{\label{multiplicity} \footnotesize Multiplicity in percentage recorded on wires for thermal neutron, gammas and fast neutron events.}
\smallskip
\begin{tabular}{|c|c|c|c|c|c|}
\hline
 & \multicolumn{5}{|c|}{Number of wires} \\
\cline{2-6}
 & 1 & 2 & 3 & 4  & 5 \\
\hline
Thermal n & 75\% & 25\% &  0\% & 0\% &  0\% \\
\hline
Gammas & 20\% & 20\% & 50\% & 10\% &  0\% \\
\hline
Fast n & 15\% & 30\% & 40\%  & 10\% & 5\% \\
\hline
\end{tabular}
\end{table}

\subsection{Fast neutron measurements with different sources}\label{sources}

\begin{figure}[htbp]
\centering
\includegraphics[width=1\textwidth,keepaspectratio]{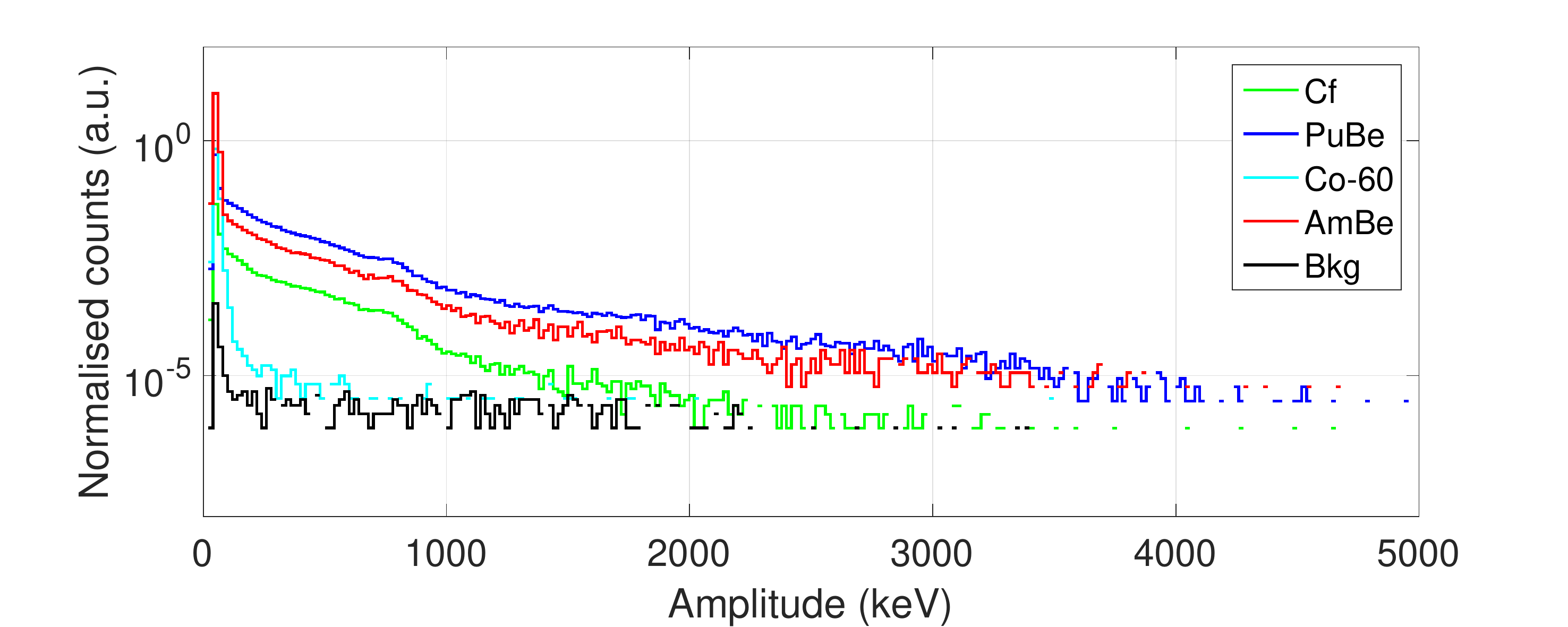}
\includegraphics[width=1\textwidth,keepaspectratio]{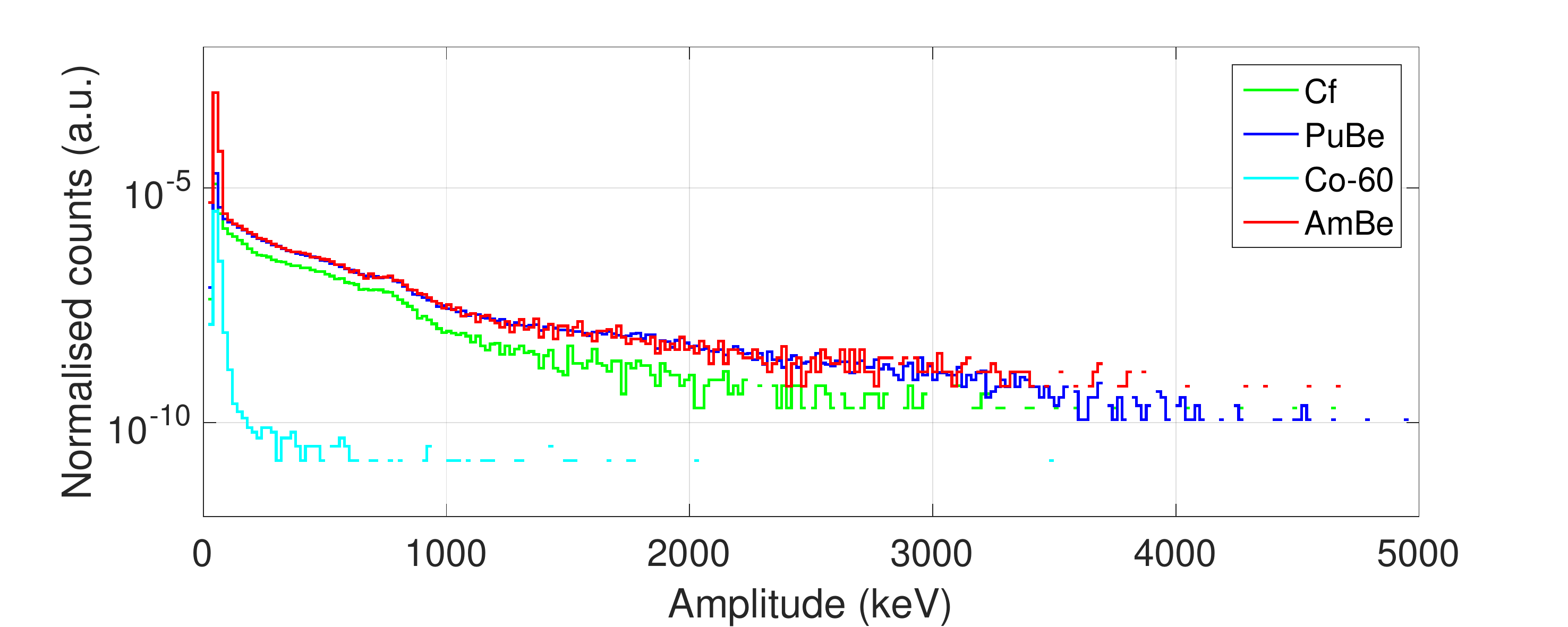}
\caption{\label{phs_sources} \footnotesize PHS of one wire for different sources and the background. Each spectrum is normalized over time (top). PHS of all the sources normalized over time, solid angle and activity (bottom).}
\end{figure}

A set of measurements have been performed with three neutron sources: $^{252}$Cf, $^{241}$Am/Be, $^{238}$Pu/Be, and with a gamma source, $^{60}$Co. The latter measurement work as a comparison with the study concerning the gamma sensitivity~\cite{MB_2017, MG_gamma}. A background measurement was performed without any sources. 
\\ These measurements provide information about the possible differences on the PHS caused by the choice of a source  (section~\ref{subsources}). While the Am/Be and Pu/Be sources spectra are similar both in shape and in energy, the distribution of a Cf source has a quite different shape in the same range of energy. Figure~\ref{phs_sources} shows the PHS for each source.

All the spectra refer to one wire, the plot at the top is normalized over the total time of acquisition for each set of measurements, while the plot at the bottom is normalized for activity and solid angle as described in subsection~\ref{subsources}. Once normalized, the PHS of the $^{241}$Am/Be, $^{238}$Pu/Be are in a good agreement because their emission energy spectra are similar both in shape and energy. The difference peak at low energy ($<$100 keV) is expected because the $^{241}$Am has a higher gamma emission ($\sim$60 keV) than that of $^{238}$Pu (figure~\ref{phs_sources}). Moreover the shape of both PHS is similar to that of Cf, which differs from the former by a factor in intensity, and a variation in shape at high energy, above 1 MeV, due to the difference between the emission energy spectrum of the sources, figure~\ref{phs_sources}. As described in section~\ref{subsources} the emission distribution of the Cf peaks at lower energy compared to the Actinide/Be sources. Both from calculations and simulations it emerges that the main contribution to the energy deposition spectrum in this region is attributed to the absorption interactions which require more energy to occur, see equations~\ref{e_recoil} and~\ref{e_y}.
%\begin{figure}[htbp]
%\centering
%\includegraphics[width=1\textwidth,keepaspectratio]{figures/source_scal_2}
%\caption{\label{phs_scale} \footnotesize PHS obtained with $^{252}$Cf scaled on the one obtained with the Actinide/Be-based radioactive sources. Less counts at higher energy due to the different shape of the emission distribution of the sources.}
%\end{figure} 
\\We can conclude that although the sources differ in energy distributions, they do not produce any relevant effect on the spectrum. Moreover, a common feature is visible around 800 keV, predicted by the theoretical calculation, figure~\ref{theo_el}, and verified by the simulations shown in figure~\ref{sim_gas_edep}. This feature can be identified as the contribution of the elastic and the absorption interactions of the gas elements to the PHS. 
\\ For the following measurements we used the $^{238}$Pu/Be source, because it has the highest activity and a lower gamma contribution among the three.

\subsection{Materials comparison}\label{materiali}

\begin{figure}[htbp]
\centering
\includegraphics[width=1\textwidth,keepaspectratio]{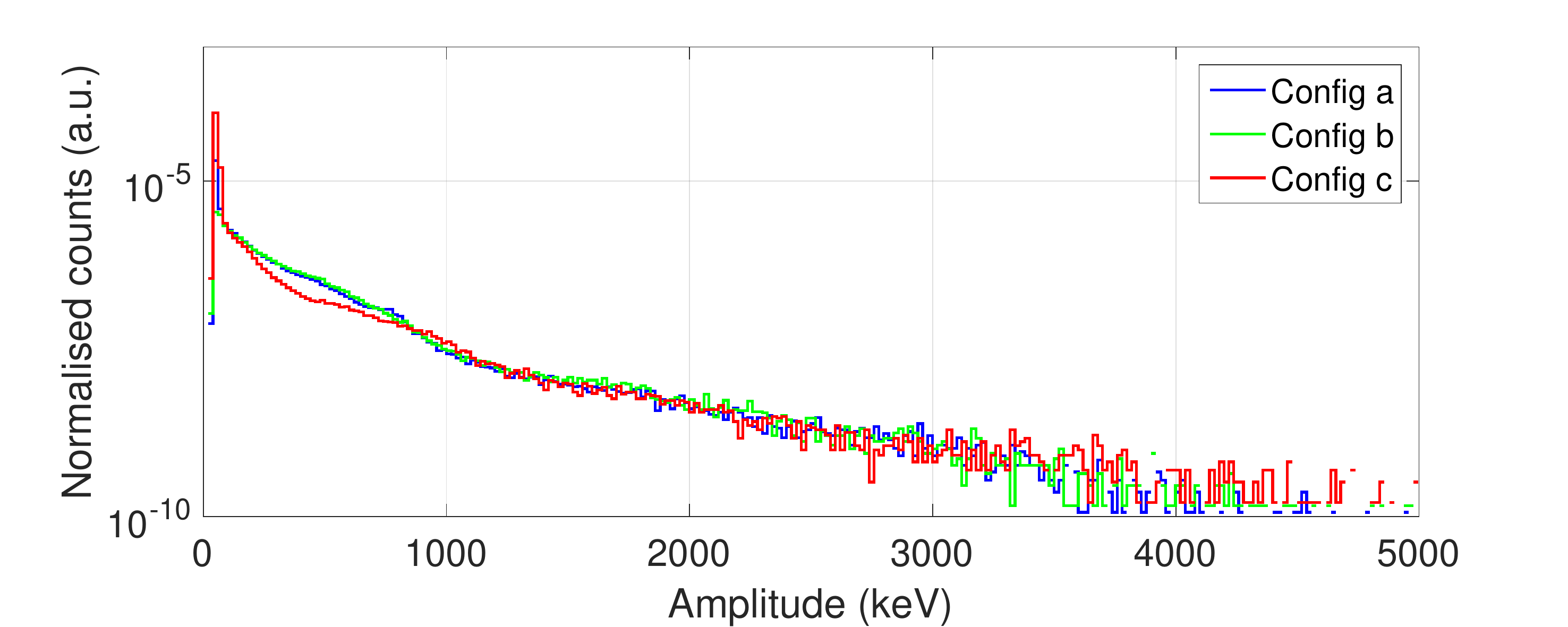}
\caption{\label{phs_material} \footnotesize The PHS are normalized by time, activity and solid angle and they refer to one single wire. The blue plot refers to the \textit{Configuration a}, the green to the \textit{Configuration b} and the red curve refers to the \textit{Configuration c} in figure~\ref{cassettesc}.}
\end{figure}

Referring to section~\ref{sp}, a significant difference between the materials of the detector is expected. We changed the cassette configuration according to figure~\ref{cassettesc}, performing the measurements for the configurations a, b and c. The PHS for each one is showed in figure~\ref{phs_material}.

From the comparison between \textit{a} and \textit{b} we can conclude that the Copper and Kapton of the strips do not contribute at our sensitivity level. The shape of the pulse height spectrum with or without the strips is similar, as expected from the theoretical calculation of the probability of deposition, see subsection~\ref{sp} and the simulations confirm it as well, see subsection~\ref{sim_contribute}. 
\\ Note that the PHS obtained with the \textit{Configuration c} (titanium blade) is different from the others below 1 MeV. The shape of the tail at higher energy is indeed unchanged between the three different setups. 
\\ A possible explanation for this behaviour can be obtained by separating solid and gas contribution to signal detection as described in section~\ref{sp}. After the neutron conversion into a charged particle, the latter has to escape and reach the gas to generate a signal. This effect in the solid is mostly significant in the low energy region of the PHS, as shown in figure~\ref{theo_el}. The gas component is the sensing medium of the detector so the conversion products release their energy directly into the gas generating a signal.
\\ From the theoretical calculation of P(r) the probability of energy deposition in the gas, see~\ref{sp} table~\ref{tabsp}, the most relevant difference is shown for the elastic scattering interaction and the (n,p) absorption reaction. The recoil nucleus of Al, from equation~\ref{al_p} and~\ref{e_y}, can release in the gas at most 690 keV for an incoming neutron energy of 5 MeV. The maximum energy released in the same condition for Ti is instead 300 keV with a probability about 4 times lower than that of Al. Moreover the P(r) of (n,p) reaction is about 1 order of magnitude higher for Al than Ti, while the energy fraction released in the gas of the emitted proton is approximately the same in both cases. The cross section of this reaction in Aluminium, as shown in figure~\ref{sigmaAlTi}, occurs at $\approx$ 3 MeV; the same reaction in Ti occurs at $\approx$ 6 MeV with roughly the same probability. The other two most significant processes take into account in the absorption interaction, namely (n,$\alpha$) and (n,$\gamma$), have approximately the same energy distribution. 
\\From this study we can attribute the difference on the PHS shape to the elastic interaction and the absorption (n,p) process. Indeed the difference is visible in the same energy range obtained from the calculation, see figure~\ref{theo_phs}, and the ratio between the two spectra is (1.5 $\pm 0.5$) in agreement with the simulations result discuss in section~\ref{sim_contribute}.  We can conclude that Titanium is less sensitive to fast neutron than Aluminium, while the background contribution of Copper and Kapton for fast neutron can be considered negligible, as expected from the theoretical considerations (subsecion ~\ref{sp}) and the simulations (subsection~\ref{sim_contribute}).
%the Copper has one of the largest microscopic cross section between the studied materials (subsection~\ref{mbdescr}) in the energy range (subsection~\ref{subsources}), but it is the haviest element among them. It is then less probable for a particle to escape the surface and release energy in the gas. 

\section{Fast neutron sensitivity}

\begin{figure}[htbp]
\centering
\includegraphics[width=1\textwidth,keepaspectratio]{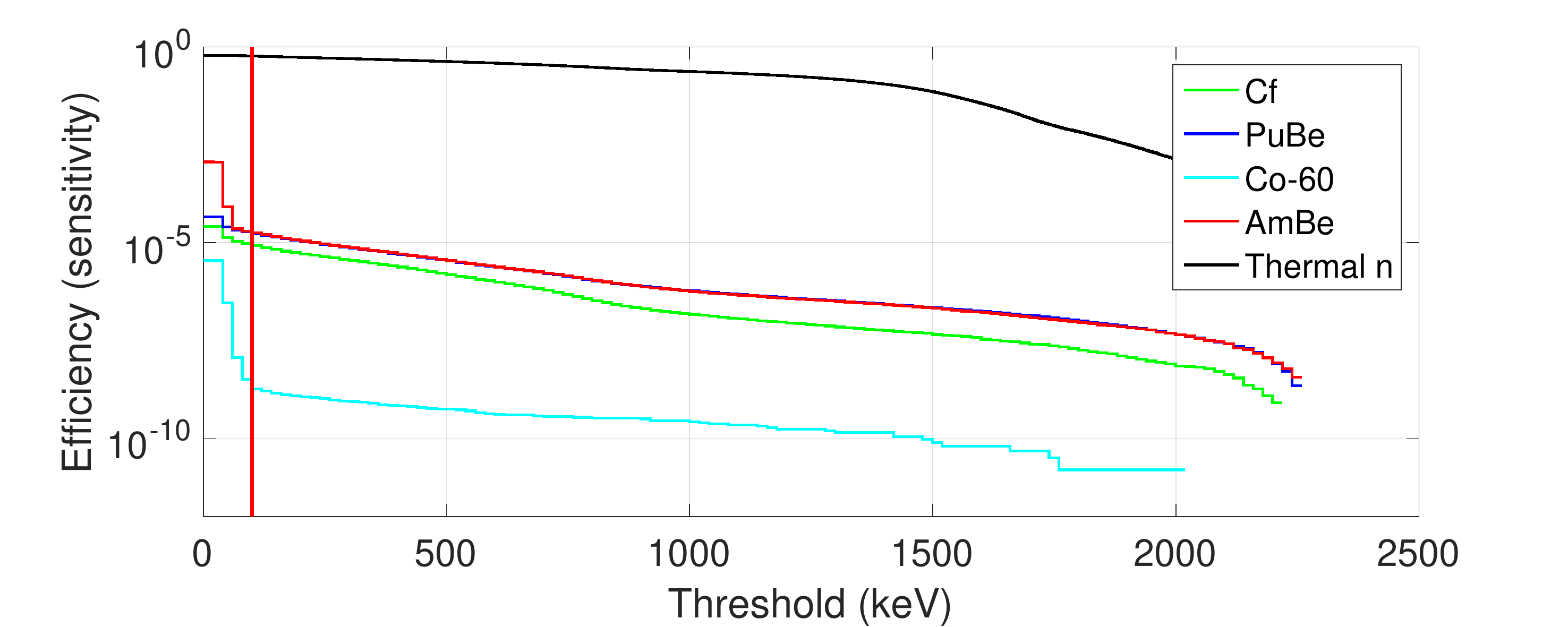}
\includegraphics[width=1\textwidth,keepaspectratio]{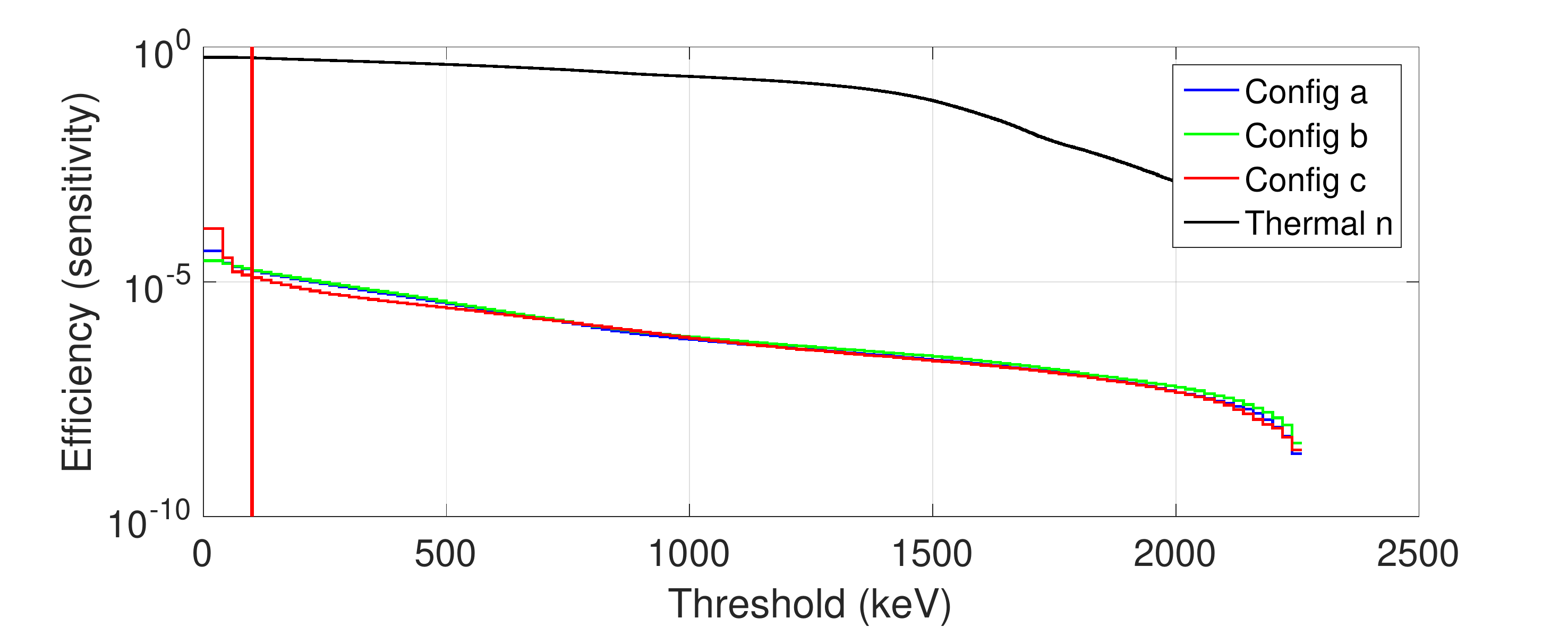}
\caption{\label{sensitivity} \footnotesize Counts in the PHS (normalized to the subtended solid angle and the activity of the sources) as a function of the applied threshold for the several sources (top). For the three configurations a, b and c (bottom).}
\end{figure}

The fast neutron sensitivity of a thermal neutron detector is defined as the probability for a fast neutron to generate a false count in a thermal/cold neutron measurement. Along with the $\gamma$-ray sensitivity~\cite{MG_gamma, MB_2017} the fast neutron sensitivity defines the best achievable signal-to-background ratio, for a given flux of each component. 
\\The interactions with fast neutrons that can give rise to background signals are described in the sections above as well as the measurements performed. When the energy for one of these processes exceed a set threshold, it results in an event. The number of events that exceed these thresholds is then normalized to the activity of the source and the solid angle. 
\\ Each source was placed close to the entrance window of the Multi-Blade, directly outside of the cassette equipped with the individual readout electronics, as showed in the sketch~\ref{setup}; the solid angle acceptance has been calculated as described in subsection~\ref{subsources}.  Note that the statistical uncertainties are almost negligible, the systematic uncertainty on the distance from the source to the sensitive element can lead to a deviation on the measurements of no more than a factor 2, subsection~\ref{subsources}.
\\ Figure~\ref{sensitivity} (top) shows the total counts in the PHS which is normalized to the activity of the source and the solid angle as a function of the threshold for all the sources used for the experiment. The bottom plot in figure~\ref{sensitivity}, shows the sensitivity for the different cassette configurations described in section~\ref{materiali}.

We report in table~\ref{fastn-gamma-thr} the values of the efficiency for thermal neutrons measured with a neutron wavelength $4.2$\AA , and for the fast neutron and gamma sources used to perform the measurements described in section~\ref{measure}. We report the values of the sensitivity at two different software threshold of 20 keV and 100 keV respectively.

\begin{table}[htbp]
\centering
\caption{\label{fastn-gamma-thr} \footnotesize Efficiency (or sensitivity) for thermal neutrons, fast neutrons and gamma-rays applying two values of the threshold (20 and 100 keV). The efficiency (or sensitivity) is normalized to 1.}
\smallskip
\begin{tabular}{|c|c|c|c|c|c|}
\hline
 & Th. n (20keV) & $^{252}$Cf (20keV) & $^{238}$Pu/Be (20keV) & $^{238}$Am/Be(20 keV)  & $^{60}$Co  (20 keV)  \\
 & \multicolumn{1}{|r|}{(100keV)} & \multicolumn{1}{|r|}{(100keV)} & \multicolumn{1}{|r|}{(100keV)} &\multicolumn{1}{|r|}{(100keV)}  &  \multicolumn{1}{|r|}{(100keV)} \\ 
\hline
Eff. & 0.60 & $2.6\cdot10^{-5}$ &  $4.5\cdot10^{-5}$ & $1.1\cdot10^{-3}$ &  $3.5\cdot10^{-6}$\\
 & 0.56 & $9.4 \cdot10^{-6}$ & $1.9\cdot10^{-5}$ & $1.9\cdot10^{-5}$ &  $<1\cdot10^{-8}$ \\
\hline
\end{tabular}
\end{table}

We fixed the threshold at 100 keV to achieve the optimal signal-to-background ratio, see vertical line in figure~\ref{sensitivity}. Indeed the efficiency calculated for the two Actinide/Be-based radioactive sources is different at low energies where the gamma contribution of $^{241}$Am/Be is more intense than that of $^{238}$Pu/Be. This effect is visible applying the software threshold of 20 keV while it is cut by the $\mathrm{100\,keV}$ threshold. The gamma sensitivity decreases by more than a factor 100 between the two values of threshold as well, see table~\ref{fastn-gamma-thr} . 
\\The efficiency for thermal neutrons measured at $4.2$\AA , is $\approx 56\%$ as described in~\cite{MB_2017}. The sensitivity at the same threshold to the fast neutron is about 4 orders of magnitude lower. Furthermore the gamma sensitivity measurement agrees with the previous dedicated studies~\cite{MG_gamma, MB_2017} and it is $\approx 10^{-8}$, about 3 order of magnitude less intense than the fast neutron sensitivity, figure~\ref{sensitivity}.
%\\ The sensitivity calculated for $^{252}$Cf source is $\sim 10^{-5}$ ($9.23 \cdot 10^{-6}$), green curve on the top plot, has a lower value than the sensitivity obtained with the $^{241}$Am/Be (red curve) $\approx 2 \cdot 10^{-5}$ and $^{238}$Pu/Be (blue curve)$\approx 2 \cdot 10^{-5}$. 
\\ The study of the various interactions that can occur between incident neutrons and the materials inside the detector is presented in this paper. Information of these reactions in Aluminium and Titanium emerges from this analysis. We focused the attention on the elastic scattering and on two most significant absorption processes, (n,p) and (n,$\alpha$). The cross section of (n,p) process in Aluminium is about 3 orders of magnitude higher than that of Titanium in the energy range of the neutron sources used for the measurements. The probability of the same mechanism in Titanium is almost out of the energy range that can be covered by the sources as discussed in section~\ref{theocalc}. 
\\ The sensitivity obtained with the Titanium blade is slightly different compared with the one calculated with the Aluminium blade at low values of amplitude, as expected by the study of the PHS discussed in section~\ref{materiali}. The sensitivity obtained with the Ti is $1.39 \cdot 10^{-5}$, the one obtained with the Al, for the same threshold ($\mathrm{100\,keV}$), is $1.88 \cdot 10^{-5}$.
\\ By choosing, for the Multi-Blade detector, blades in Titanium instead of blades in Aluminium, it is possible to improve the signal-to-background ratio of about a factor two, at the fixed software threshold of 100 keV. This change also brings mechanical advantage, as reported in~\cite{MB_2017}.

\section{Conclusions and outlook}

The fast neutron sensitivity in many types of thermal neutron detector is an important feature to characterize, to achieve the best signal-to-background ratio. As the search for alternatives for the $^3$He crisis began, it is interesting to investigate the performances of the new detector technologies. In particular this work focuses the attention on which fast neutron sensitivity can be reached with the new technologies. 
\\ In the paper we have investigated the physical effects that affect the fast neutron sensitivity in Boron-10-based gaseous detectors for thermal neutrons which are being developed at ESS. A full characterization of detectors is necessary to fulfil the ESS requirements of high performance, the background suppression ($\gamma$-ray, fast neutron sensitivity) as well. All the measurements presented were done for the Multi-Blade detector and performed at the Source Testing Facility (STF) at Lund University in Sweden.

The investigation of the physical mechanisms, cross section and the theoretical calculation of the probability of interaction in several materials, together with the simulations of the same processes provide an overview and an overall understanding of the work presented in the paper.

The fast neutron sensitivity of the Multi-Blade has been measured for three fast neutron sources ($^{252}$Cf, $^{241}$Am/Be, $^{238}$Pu/Be) and it was found that for a fixed software threshold of 100 keV, which leads to the optimal neutron efficiency, it is approximately 10$^{-5}$, with an uncertainty no more than a factor two dominated by the calculation of the solid angle. A comparison between the $\gamma$-ray sensitivity and the fast neutron sensitivity has been performed. We used for the $\gamma$-ray measurements a $^{60}$Co source. We obtained a value for the $\gamma$-ray sensitivity (below 10$^{-8}$) about 3 orders of magnitude less intense than the fast neutron sensitivity for the same threshold. The value for gamma sensitivity comes from both a low interaction cross section and from a low probability of detection of a signal over a threshold. For fast neutron sensitivity, only the low interaction probability helps keep the sensitivity low.

The measurements presented here were performed for a thermal neutron detector for the first time. A $^{10}$B-based gaseous detector was chosen. The Multi-Blade is based on a MWPC geometry, the results obtained and the discussion of the underlying physical processes responsible for fast neutron events in a neutron detector are more general and can be extended other gaseous-based neutron detectors. 

\acknowledgments This work is being supported by the BrightnESS project (Horizon 2020, INFRADEV-3-2015, grant number 676548) and carried out as a part of the collaboration between the European Spallation Source (ESS - Sweden), the Lund University (LU - Sweden), the Link\"{o}ping University (LiU - Sweden), the Wigner Research Centre for Physics (Hungary) and the University of Perugia (Italy).

The work was carried out at the Source Testing Facility, Lund University (LU - Sweden).\\
Computing resources provided by DMSC Computing Centre:\\ \url{https://europeanspallationsource.se/data-management-software/computing-centre}.

\bibliographystyle{ieeetr}
\bibliography{BIBLIODB}

\begin{thebibliography}{10}

\bibitem{ESS}
European Spallation Source ESS ERIC - http://europeanspallationsource.se.

\bibitem{ESS_TDR}
S.~Peggs. ESS Technical Design Report - (ESS-2013-0001).

\bibitem{HE3S_karl}
K.~Zeitelhack, ``Search for alternative techniques to Helium-3 based detectors
  for neutron scattering applications,'' {\em Neutron News}, vol.~23, no.~4,
  pp.~10--13, 2012.

\bibitem{MG_2017}
M.~Anastasopoulos, R.~Bebb, K.~Berry, J.~Birch, T.~Bry{\'s}, J.-C. Buffet,
  J.-F. Clergeau, P.~Deen, G.~Ehlers, P.~van Esch, S.~Everett, B.~Guerard,
  R.~Hall-Wilton, K.~Herwig, L.~Hultman, C.~H{\"o}glund, I.~Iruretagoiena,
  F.~Issa, J.~Jensen, A.~Khaplanov, O.~Kirstein, I.~L. Higuera, F.~Piscitelli,
  L.~Robinson, S.~Schmidt, and I.~Stefanescu, ``Multi-Grid detector for neutron
  spectroscopy: results obtained on time-of-flight spectrometer CNCS,'' {\em
  Journal of Instrumentation}, vol.~12, no.~04, p.~P04030, 2017.

\bibitem{MG_IN6tests}
J.~Birch, J.-C. Buffet, J.-F. Clergeau, J.~Correa, P.~van Esch, M.~Ferraton,
  B.~Guerard, J.~Halbwachs, R.~Hall-Wilton, L.~Hultman, C.~H{\"o}glund,
  A.~Khaplanov, M.~Koza, F.~Piscitelli, and M.~Zbiri, ``In-beam test of the
  boron-10 Multi-Grid neutron detector at the IN6 time-of-flight spectrometer
  at the ill,'' {\em Journal of Physics: Conference Series}, vol.~528, no.~1,
  p.~012040, 2014.

\bibitem{MG_patent}
B.~Guerard and J.~Buffet, ``Ionizing radiation detector - patent no.
  20110215251.,'' Sept.~8 2011.
\newblock US Patent App. 13/038,915.

\bibitem{MG_joni}
J.~Birch, J.~C. Buffet, J.~Correa, P.~van Esch, B.~Gu{\'e}rard, R.~Hall-Wilton,
  C.~H{\"o}glund, L.~Hultman, A.~Khaplanov, and F.~Piscitelli, ``10B4C
  Multi-Grid as an alternative to 3He for large area neutron detectors,'' {\em
  IEEE Transactions on Nuclear Science}, vol.~60, pp.~871--878, April 2013.

\bibitem{MIO_MB2014}
F.~Piscitelli, J.~C. Buffet, J.~F. Clergeau, S.~Cuccaro, B.~Gu{\'e}rard,
  A.~Khaplanov, Q.~L. Manna, J.~M. Rigal, and P.~V. Esch, ``Study of a high
  spatial resolution 10B-based thermal neutron detector for application in
  neutron reflectometry: the Multi-Blade prototype,'' {\em Journal of
  Instrumentation}, vol.~9, no.~03, p.~P03007, 2014.

\bibitem{MB_2017}
F.~Piscitelli, F.~Messi, M.~Anastasopoulos, T.~Bry{\'s}, F.~Chicken, E.~Dian,
  J.~Fuzi, C.~H{\"o}glund, G.~Kiss, J.~Orban, P.~Pazmandi, L.~Robinson,
  L.~Rosta, S.~Schmidt, D.~Varga, T.~Zsiros, and R.~Hall-Wilton, ``The
  Multi-Blade Boron-10-based neutron detector for high intensity neutron
  reflectometry at ESS,'' {\em Journal of Instrumentation}, vol.~12, no.~03,
  p.~P03013, 2017.

\bibitem{DET_jalousie}
M.~Henske, M.~Klein, M.~K{\"o}hli, P.~Lennert, G.~Modzel, C.~Schmidt, and
  U.~Schmidt, ``The 10B-based jalousie neutron detector -- an alternative for
  3He filled position sensitive counter tubes,'' {\em Nuclear Instruments and
  Methods in Physics Research Section A: Accelerators, Spectrometers, Detectors
  and Associated Equipment}, vol.~686, pp.~151 -- 155, 2012.

\bibitem{MPGD_GEMcroci}
G.~Croci and et~al., ``Diffraction measurements with a boron-based {GEM}
  neutron detector,'' {\em EPL}, vol.~107, no.~1, p.~12001, 2014.

\bibitem{Bgem}
G.~Albani, E.~P. Cippo, G.~Croci, A.~Muraro, E.~Schooneveld, A.~Scherillo,
  R.~Hall-Wilton, K.~Kanaki, C.~H{\"o}glund, L.~Hultman, J.~Birch, G.~Claps,
  F.~Murtas, M.~Rebai, M.~Tardocchi, and G.~Gorini, ``Evolution in boron-based
  gem detectors for diffraction measurements: from planar to 3D converters,''
  {\em Measurement Science and Technology}, vol.~27, no.~11, p.~115902, 2016.

\bibitem{STRAW_lacy2011}
J.~L. Lacy, A.~Athanasiades, L.~Sun, C.~S. Martin, T.~D. Lyons, M.~A. Foss, and
  H.~B. Haygood, ``Boron-coated straws as a replacement for 3He-based neutron
  detectors,'' {\em Nuclear Instruments and Methods in Physics Research Section
  A: Accelerators, Spectrometers, Detectors and Associated Equipment},
  vol.~652, no.~1, pp.~359 -- 363, 2011.
\newblock Symposium on Radiation Measurements and Applications (SORMA) \{XII\}
  2010.

\bibitem{DET_doro1}
D.~Pfeiffer, F.~Resnati, J.~Birch, M.~Etxegarai, R.~Hall-Wilton,
  C.~H{\"o}glund, L.~Hultman, I.~Llamas-Jansa, E.~Oliveri, E.~Oksanen,
  L.~Robinson, L.~Ropelewski, S.~Schmidt, C.~Streli, and P.~Thuiner, ``First
  measurements with new high-resolution gadolinium-{GEM} neutron detectors,''
  {\em Journal of Instrumentation}, vol.~11, no.~05, p.~P05011, 2016.

\bibitem{gdgem}
D.~Pfeiffer, F.~Resnati, J.~Birch, R.~Hall-Wilton, C.~H{\"o}glund, L.~Hultman,
  G.~Iakovidis, E.~Oliveri, E.~Oksanen, L.~Ropelewski, and P.~Thuiner, ``The
  u-TPC method: improving the position resolution of neutron detectors based on
  mpgds,'' {\em Journal of Instrumentation}, vol.~10, no.~04, p.~P04004, 2015.

\bibitem{DET_kohli}
M.~K{\"o}hli, F.~Allmendinger, W.~H{\"a}u{\ss}ler, T.~Schr{\"o}der, M.~Klein,
  M.~Meven, and U.~Schmidt, ``Efficiency and spatial resolution of the cascade
  thermal neutron detector,'' {\em Nuclear Instruments and Methods in Physics
  Research Section A: Accelerators, Spectrometers, Detectors and Associated
  Equipment}, vol.~828, pp.~242--249, 8 2016.

\bibitem{IAEA-STF}
F.~Messi, H.~Perrey, K.~Fissum, M.~Akkawi, R.~A. Jebali, J.~R.~M. Annand, P.~M.
  Bentley, L.~Boyd, C.~P. Cooper-Jensen, D.~D. DiJulio, J.~Freita-Ramos,
  R.~Hall-Wilton, A.~Huusko, T.~Ilves, F.~Issa, A.~Jalg\'en, K.~Kanaki,
  E.~Karnickis, A.~Khaplanov, S.~Koufigar, V.~Maulerova, G.~Mauri,
  N.~Mauritzson, W.~Pei, F.~Piscitelli, E.~Rofors, J.~Scherzinger,
  H.~S\"oderhielm, D.~S\"oderstr\"om, and I.~Stefanescu, ``The neutron tagging
  facility at Lund university,'' 2017. arXiv: 1711.10286.

\bibitem{MG_gamma}
A.~Khaplanov, F.~Piscitelli, J.~C. Buffet, J.~F. Clergeau, J.~Correa, P.~van
  Esch, M.~Ferraton, B.~Guerard, and R.~Hall-Wilton, ``Investigation of
  gamma-ray sensitivity of neutron detectors based on thin converter films,''
  {\em Journal of Instrumentation}, vol.~8, no.~10, p.~P10025, 2013.

\bibitem{MPGD_CrociGamma}
G.~Croci, C.~Cazzaniga, M.~Tardocchi, R.~Borghi, G.~Claps, G.~Grosso,
  F.~Murtas, and G.~Gorini, ``Measurements of $\gamma$-ray sensitivity of a
  {GEM} based detector using a coincidence technique,'' {\em Journal of
  Instrumentation}, vol.~8, no.~04, p.~P04006, 2013.

\bibitem{B4C_carina}
C.~H{\"o}glund, J.~Birch, K.~Andersen, T.~Bigault, J.-C. Buffet, J.~Correa,
  P.~van Esch, B.~Guerard, R.~Hall-Wilton, J.~Jensen, A.~Khaplanov,
  F.~Piscitelli, C.~Vettier, W.~Vollenberg, and L.~Hultman, ``B4C thin films
  for neutron detection,'' {\em Journal of Applied Physics}, vol.~111, no.~10,
  2012.

\bibitem{B4C_Schmidt}
S.~Schmidt, C.~H{\"o}glund, J.~Jensen, L.~Hultman, J.~Birch, and
  R.~Hall-Wilton, ``Low-temperature growth of boron carbide coatings by direct
  current magnetron sputtering and high-power impulse magnetron sputtering,''
  {\em Journal of Materials Science}, vol.~51, no.~23, pp.~10418--10428, 2016.

\bibitem{MIO_analyt}
F.~Piscitelli and P.~V. Esch, ``Analytical modelling of thin film neutron
  converters and its application to thermal neutron gas detectors,'' {\em
  Journal of Instrumentation}, vol.~8, no.~04, p.~P04020, 2013.

\bibitem{EL_cremat}
CREMAT Inc - Electronics for pulse detection - http://www.cremat.com.

\bibitem{EL_CAEN}
CAEN - Electronic Instrumentation - http://www.caen.it.

\bibitem{PuBe_AmBe}
J.~Scherzinger, R.~A. Jebali, J.~Annand, K.~Fissum, R.~Hall-Wilton,
  S.~Koufigar, N.~Mauritzson, F.~Messi, H.~Perrey, and E.~Rofors, ``A
  comparison of untagged gamma-ray and tagged-neutron yields from 241AmBe and
  238PuBe sources,'' {\em Applied Radiation and Isotopes}, vol.~127, pp.~98 --
  102, 2017.

\bibitem{Cf}
J.~Scherzinger, R.~Jebali, J.~Annand, A.~Bala, K.~Fissum, R.~Hall-Wilton,
  D.~Hamilton, N.~Mauritzson, F.~Messi, H.~Perrey, and E.~Rofors, ``Tagging
  fast neutrons from a 252Cf fission-fragment source,'' {\em Applied Radiation
  and Isotopes}, pp.~--, 2017.

\bibitem{DET_knoll}
G.~Knoll, {\em Radiation Detection and Measurement}.
\newblock John Wiley and Sons, Inc., third edition~ed., 2000.

\bibitem{n_interaction_matter}
R.~P., ``Neutron interactions with matter,'' tech. rep., Los Alamos Technical
  Report, http://www.fas.org/sgp/othergov/doe/
  lanl/lib-www/la-pubs/00326407.pdf, 2009.

\bibitem{NIST}
NIST National Nuclear Data Center - https://www.nndc.bnl.gov.

\bibitem{Al_np}
``Measurement of neutron activation cross sections in the energy range between
  2 and 7 MeV by using a Ti-deuteron target and a deuteron gas target,'' {\em
  Proceedings of the JAERI Conference 2000, Japan,}, pp.~208--213, 2000.

\bibitem{Ti_np}
N.~Molla and S.~Qaim, ``A systematic study of (n, p) reactions at 14.7 MeV,''
  {\em Nuclear Physics A}, vol.~283, no.~2, pp.~269 -- 288, 1977.

\bibitem{O_nalpha}
D.~J. VEAL and C.~F. COOK, ``A rapid method for the direct determination of
  elemental oxygen by activation with fast neutrons,'' {\em Anal. Chem.},
  vol.~34, pp.~178--184, February 1962.

\bibitem{MISC_SRIM2010}
J.~F. {Ziegler}, M.~D. {Ziegler}, and J.~P. {Biersack}, ``{SRIM - The stopping
  and range of ions in matter (2010)},'' {\em Nuclear Instruments and Methods
  in Physics Research B}, vol.~268, pp.~1818--1823, June 2010.

\bibitem{MISC_SRIM1998}
J.~F. {Ziegler}, ``{RBS/ERD simulation problems: Stopping powers, nuclear
  reactions and detector resolution},'' {\em Nuclear Instruments and Methods in
  Physics Research B}, vol.~136, pp.~141--146, Mar. 1998.

\bibitem{g1}
{S. Agostinelli et al.}, ``{GEANT4 ---- a simulation toolkit},'' {\em Nuclear
  Instruments and Methods in Physics Research A}, vol.~506, pp.~250--303, 2003.

\bibitem{g2}
{J. Allison et al.}, ``{GEANT4 developments and applications},'' {\em IEEE
  Transactions on Nuclear Science}, vol.~53, pp.~270--278, 2006.

\bibitem{g3}
{J. Allison et al.}, ``{Recent developments in GEANT4},'' {\em Nuclear
  Instruments and Methods in Physics Research A}, vol.~835, pp.~186--225, 2016.

\bibitem{ncrystal}
``{NCrystal: a library for thermal neutron transport in crystals}.''
  \url{https://github.com/mctools/ncrystal/wiki}
  https://doi.org/10.5281/zenodo.853186, 2017.

\end{thebibliography}
\end{document}